\documentclass[preprint,12pt]{elsarticle}

\usepackage{amssymb}
\usepackage{amsmath}
\usepackage{xcolor}
\usepackage{graphicx}
\usepackage{siunitx}
\usepackage{enumitem}
\usepackage{ulem}
\usepackage[margin=0.7in]{geometry}

\usepackage{hyperref}
\usepackage{mathrsfs}

\usepackage[section]{placeins} % Place figures in the section where they were introduced

\usepackage{datetime} % Tweak date format. I don't like showing day ...

\usepackage[numbers]{natbib}

\usepackage[most]{tcolorbox}

\usepackage{multirow}

\usepackage[nameinlink]{cleveref}
\Crefname{equation}{Equation}{Equations} 
\crefname{equation}{}{} 
\crefname{subsection}{subsection}{subsections}
\crefname{appendix}{}{}

\renewcommand{\[}{\left[}
\renewcommand{\]}{\right]}
\renewcommand{\(}{\left(}
\renewcommand{\)}{\right)}

\newcommand{\dd}{{\rm d}}

\newcommand{\fracp}[2]{\frac{\partial #1}{\partial #2}}

\newcommand{\bB}{{\bf{B}}}

\newcommand{\bF}{{\bf{F}}}

\newcommand{\bH}{{\bf{H}}}
\newcommand{\bI}{{\bf{I}}}

\newcommand{\bM}{{\bf{M}}}

\newcommand{\bP}{{\bf{P}}}

\newcommand{\bX}{{\boldsymbol{X}}}

\newcommand{\bb}{{\boldsymbol{b}}}

\newcommand{\bn}{{\boldsymbol{n}}}

\newcommand{\bt}{{\boldsymbol{t}}}
\newcommand{\bu}{{\boldsymbol{u}}}
\newcommand{\bv}{{\boldsymbol{v}}}
\newcommand{\bw}{{\boldsymbol{w}}}
\newcommand{\bx}{{\boldsymbol{x}}}

\newcommand{\beps}{{\boldsymbol{\epsilon}}}

\newcommand{\bsig}{{\boldsymbol{\sigma}}}

\newcommand{\bvphi}{{\boldsymbol{\varphi}}}

\newcommand{\bzero}{{\bf{0}}}

\definecolor{okabe-ito_orange}{RGB}{230,159,0}
\definecolor{okabe-ito_lblue}{RGB}{86,180,233}
\definecolor{okabe-ito_dblue}{RGB}{0,114,178}
\definecolor{okabe-ito_green}{RGB}{0,158,115}
\definecolor{okabe-ito_red}{RGB}{213,94,0}
\definecolor{okabe-ito_pink}{RGB}{204,121,167}

\newdateformat{monthyear}{\monthname[\THEMONTH] \THEYEAR}

\newcommand{\basis}{\boldsymbol{e}}

\newcommand{\refX}{\mathscr{X}}
\newcommand{\refY}{\mathscr{Y}}
\newcommand{\refZ}{\mathscr{Z}}
\newcommand{\refZbar}{\overline{\refZ}}

\newcommand{\Gbar}{\overline{G}}
\newcommand{\Ibar}{\bar{I}}
\newcommand{\Zbar}{\overline{Z}}

\newcommand{\blob}{\mathcal{C}}
\newcommand{\driveQ}{\blob_1}
\newcommand{\stiffK}{\blob_2}
\newcommand{\altQ}{\blob_3}
\newcommand{\altK}{\blob_4}

\newcommand{\Htwo}{\mathcal{H}}
\newcommand{\HL}{\Upsilon}

\newcommand{\tr}[1]{ {\rm tr}\( #1 \) }

\newcommand{\Grad}[1]{ {\rm Grad}\( #1 \) }
\newcommand{\Div}[1]{ {\rm Div}\( #1 \) }
\newcommand{\Lapl}[1]{ {\Div{\Grad{ #1 }}} }

\newcommand{\bigv}{\Big{|}}

\newcommand{\bigo}{{O}}

\newcommand{\transp}{{\top}}    % Transpose

\newcommand{\sign}{{\rm sign}}    % Transpose

\newcommand{\iden}{\bI}
\newcommand{\defgrad}{\bF}
\newcommand{\onegrad}{\bH}
\newcommand{\twograd}{\bM}

\newcommand{\defmap}{\bvphi}

\newcommand{\shifter}{\mathbf{S}}

\newcommand{\leftCG}{\bB}
\newcommand{\cauchy}{\bsig}
\newcommand{\pkone}{\bP}

\journal{Elsevier Journal TBD}

\begin{document}

\begin{frontmatter}



\title{Migration of inflated cavities in graded hyperelastic solids} %% Article title


\author[um_me]{Zhiren Zhu} 
\author[um_me]{Jonathan B. Estrada} 

\affiliation{organization={University of Michigan},%Department and Organization
            addressline={2350 Hayward Street}, 
            city={Ann Arbor},
            % postcode={48109}, 
            state={Michigan},
            country={USA}}

%% Abstract
\begin{abstract}
The inflation of a pre-existing, fluid-filled cavity is a timeless topic in the finite-deformation analysis of soft materials.
However, classical solutions for cavity inflation rely on radially symmetric material properties, leaving unresolved the effects of non-radial stiffness heterogeneity that are commonly present in biological tissues and engineered soft materials.
In this work, we investigate the quasi-static inflation of a pressurized cavity in a hyperelastic solid with shear modulus varying monotonically along a reference Cartesian direction. 
Finite-element simulations reveal that, beyond an initial small-inflation regime, the most pronounced symmetry-breaking response is the migration of the cavity toward the more compliant end of the material, while nonspherical distortion remains comparatively weak. 
To analytically quantify this migration-dominated response, we develop a Rayleigh--Ritz reduced-order framework to determine the strain-energy-minimizing migration amplitude for prescribed gradation parameters and inflation level.
Without using fitted parameters, the Rayleigh--Ritz framework recovers key features of the cavity migration that are intimately linked to the mechanical gradation parameters.
The identification of centroid migration as a salient geometric signal, together with the reduced-order prediction of its evolution, suggests a roadmap for inverse characterization of graded materials through cavity-inflation experiments.
\end{abstract}


% \begin{highlights}
% \end{highlights}

%% Keywords

\end{frontmatter}


%% NOTES AT TOP
% \zz{
% \emph{Status update (7/13/26)}:
% Updated text and dumped placeholder figures.
% Figures will be beautified after discussion with Jon ...
% }

%% ==================================
%% MAIN TEXT ========================
\section{Introduction}
\label{sec:intro}

The inflation of a pressurized cavity in a soft solid is a classical problem that remains relevant to modern challenges in engineering and biomedicine. 
The poker-chip experiments by \citet{gent-lindley_1957_internal-flaws, gent-lindley_1959_internal-rupture} showed that, when a vulcanized rubber specimen is subjected to a stiffness-dependent critical load, internal cavities may spontaneously grow. 
This phenomenon, known as cavitation, is considered to be a precursor to~\cite{gent-lindley_1959_internal-rupture,gent1990cavitation}, or a companion of~\cite{raayai-ardakani-etal_2019_cavtation-and-fracture, breedlove-etal_2024_cavitation-review}, fracture in elastomers.
Following breakthroughs in the fabrication of soft, multi-material and architected structures~\cite{truby-lewis_2016_soft-printing,holzl-etal_2016_review-bioink,hospodiuk-etal_2017_review-bioink}, there is now an increased need to assess the mechanical roles of pre-existing cavities, present in the form of designed porosity, process defects, entrapped air, etc.
In recent years, needle-~\cite{zimberlin-etal_2007_cavitation-rheometry,raayai-ardakani-cohen_2019_vcce,chockalingam-etal_2021_vcce} and laser-induced~\cite{estrada-etal_2018_IMR,zhu-etal_2025_pIMR,kolluri_2025_thin-IMR} cavitation experiments have gained popularity as a minimally-destructive approach to assess the local properties of hydrogels and tissues.
The excitation of microbubbles in soft biomaterials is also the basis of minimally-invasive surgical procedures~\cite{xu-etal_2024_review-histotripsy, deSaintVictor-etal_2014_sonothrombolysis} and targeted drug delivery methods~\cite{coussios-roy_2008_review-acoustic-therapy-drug-delivery, chowdhury-etal_2020_ultrasound-microbubble-delivery, abeid-etal_2024_droplet-cavitation}.
The ability to characterize and predict the interaction of the cavity and its hyperelastic surrounding is fundamental for these applications.

In many cases, soft materials exhibit mechanical gradation. 
A smooth spatial transition of stiffness is observed in biomaterials such as squid beaks~\cite{miserez-etal_2008_squid-beak} and sucker ring teeth~\cite{miserez-etal_2009_squid-sucker-ring}, byssal threads of marine mussels~\cite{coyne1997extensible,waite-etal_2002_marine-holdfasts}, and entheses connecting bones to tendons and ligaments in human shoulder and knee joints~\cite{benjamin-etal_2006_attachment-sites, lu-thomopoulos_2013_functional-attachment}.
Commonly achieved through the transition of chemical constituents or nanopore fraction, the gradual change of stiffness reduces stress concentration at the interface of dissimilar materials~\cite{suresh_2001_graded, liu-etal_2017_review-graded-biomaterial}.
These biological systems have also inspired recent advances in the fabrication of elastomeric composites and hydrogels with stiffness gradation~\cite{wan-etal_2024_4d-print, li-etal_2021_gradient-biomaterials, pragya-ghosh_2023_soft-graded-review}.

How does cavity inflation proceed in a hyperelastic material with stiffness gradation?
To our knowledge, this question remains largely unexplored.
In the classical context of a homogeneous, isotropic, incompressible material, the inflation of an initially spherical cavity results in an analytically tractable displacement field that preserves spherical symmetry---i.e., material points only move radially~\cite{ball_1982_cavitation, horgan-polignone_1995_cavitation-review}.
For a heterogeneous material, analytical solutions are still available if mechanical properties vary only in the radial direction from the cavity center, thereby retaining spherical symmetry~\cite{horgan-pence_1989_composite-cavity, polignone-horgan_1993_composite-aniso-cav, mousavi2023analysis}.
Yet, in most practical cases, we cannot guarantee this form of idealized alignment between the cavity and the material heterogeneity.

The present work examines such a misalignment by considering the inflation of a cavity in a hyperelastic body with stiffness gradation along a reference Cartesian direction.
We formulate the problem in \Cref{sec:formulation}.
In \Cref{sec:fe-results}, we present finite-element simulations examining the effects of the stiffness gradation on the inflated cavity geometry. 
The simulations reveal that, beyond an initial small-inflation regime, the cavity's spherical symmetry is broken primarily in the form of centroid migration, rather than pronounced nonspherical distortion.
To elucidate the underlying mechanics behind the migration-dominated response, we introduce, in \Cref{sec:first-order}, a Rayleigh--Ritz reduced-order model to analytically solve for the strain-energy-minimizing migration amplitude.
The reduced-order model is extended in \Cref{sec:second-order} to account for geometric nonlinearities associated with finite deformation, resulting in an improved agreement with the finite-element results over a broad range of gradation length scales, stiffness contrasts, and inflation levels.
In \Cref{sec:discussion}, we assess the limitations of the Rayleigh--Ritz predictions and highlight the broader utility of the proposed reduced-order framework.
Notably, we outline a roadmap to recover the mechanical gradation parameters from a sequence of quasi-static cavity inflation experiments.

% =====================
\section{Problem formulation}
\label{sec:formulation}

\begin{figure}[tb]
    \centering
    \includegraphics[width=0.75\linewidth]{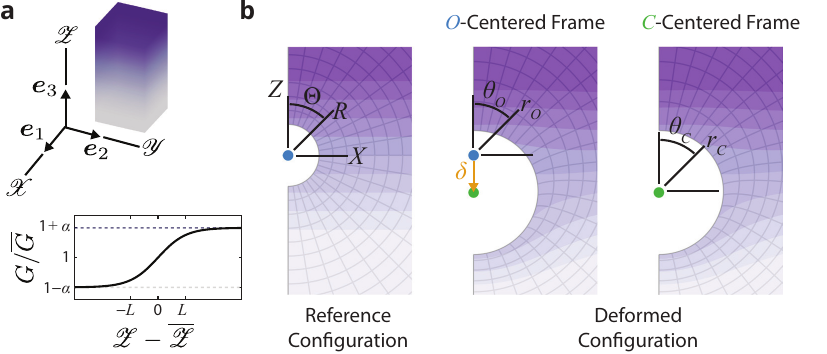}
    \caption{
    Problem geometry, stiffness gradation, and coordinate systems.
    (a) The specimen is idealized as an unbounded body of graded material, with shear modulus increasing monotonically in the $\basis_3$ direction.  
    (b) The pressurized cavity and its surrounding environment, viewed in the $\phi = \Phi = 0$ meridian plane: 
    in the reference configuration, the center of the cavity is positioned at $O$, which is chosen as the origin of the cavity-centered coordinate frame in our analysis;
    in the deformed configuration, the cavity inflates to a volume of $V = 4 \, \pi \, a_V^3/3$ and its centroid migrates to $C$.
    Modal decomposition of the cavity geometry is performed in the $C$-centered coordinate frame.
    }
    \label{fig:geometry-schematics}
\end{figure}

\subsection{Geometry and material gradation}

We consider a specimen idealized as an unbounded body.
In the reference configuration, we introduce a specimen-fixed Cartesian frame with coordinates $\{\refX, \refY, \refZ\}$ corresponding to orthonormal basis vectors $\{\basis_1, \basis_2, \basis_3\}$, as illustrated in \Cref{fig:geometry-schematics}~(a). 
The basis vectors are fixed in the ambient Euclidean space and used for both the reference and deformed configurations in our analysis.
The unbounded body is composed of an incompressible, hyperelastic material with local shear modulus $G$ monotonically graded along the direction of $\basis_3$.
In other words, the material has constant shear modulus along planes with constant $\refZ$. 
Because the shear modulus $G$ is the only spatially varying constitutive parameter, we refer to its spatial variation as the stiffness gradation. 

In our analysis, we consider the stiffness gradation profile, 
\begin{equation}
	G\( \refZ \) = \Gbar \( 1 + \alpha \tanh\( \frac{\refZ - \refZbar}{L} \) \) 
\, ,
\label{eqn:gradation_ref}
\end{equation}
where $\Gbar$ is the mean shear modulus, $\refZbar$ specifies the location of the gradation midplane (corresponding to $G = \Gbar$), $L$ is a characteristic length scale of the gradation, and $\alpha$ is the gradation amplitude.
The gradation amplitude may be expressed as
\begin{equation}
	\alpha = \frac{\Gamma-1}{\Gamma + 1}
\, ,
\end{equation}
where
\begin{equation}
    \Gamma = \frac{G\(\refZ \to +\infty\)}{G\(\refZ \to -\infty \)}
\, 
\end{equation}
is the stiffness contrast in the graded material.
For example, if $\Gamma = 2$, then the shear modulus at the stiffest end of the material is twice the shear modulus at the most compliant end of the material. 
Henceforth, we assume that $\Gamma > 1$ and the shear modulus $G$ increases with $\refZ$.

The functional form of \cref{eqn:gradation_ref} is chosen as a compact representation of a monotonic gradation profile with bounded maximum and minimum values. 
As illustrated in \Cref{fig:geometry-schematics}~(a), at the $\refZ \to -\infty$ and $\refZ \to +\infty$ limits, we have plateaus of homogeneous shear moduli that differ by a factor of $\Gamma$.
The smooth transition of the shear modulus between the compliant and stiff plateaus is described by a single length scale $L$.
In the limit where $L \to 0$, the gradation profile corresponds to a sharp bi-material interface at $\refZ = \refZbar$ and $\alpha$ is equivalent to Dundurs’ first parameter, describing the shear modulus mismatch at the interface~\cite{dundurs_1969_edge-bonded-wedges, hutchinson-suo_1991_mixed-mode-crack}.
While the gradation profile described by \cref{eqn:gradation_ref} offers analytical convenience, our analysis procedure is not restricted to this specific functional form.
It can be adapted to consider other sufficiently smooth, monotonic profiles featuring limiting moduli and transition length scales.

In the reference configuration, the unbounded body is stress-free and contains a spherical cavity with undeformed radius $A$.
Let $O$, with coordinates $\{\refX_O, \refY_O, \refZ_O\}$ in the specimen-fixed frame, denote the fixed spatial point coinciding with the center of the undeformed cavity.
To facilitate our analysis, we introduce a cavity-centered reference frame, as illustrated in \Cref{fig:geometry-schematics}~(b), with material points described by its position relative to $O$,
\begin{equation}
	X = \refX - \refX_O
\, , \quad
	Y = \refY - \refY_O
\, , \quad
	Z = \refZ - \refZ_O
\, ,
\end{equation}
or using an associated set of spherical coordinates $\{R, \Theta, \Phi\}$ satisfying 
\begin{equation}
    R = \(X^2 + Y^2 + Z^2\)^{1/2}
    \, , \quad
    \cos\Theta = Z/R
    \, , \quad
    \tan\Phi = Y/X
    \, .
\end{equation}

In the cavity-centered frame, the stiffness gradation profile is described as 
\begin{equation}
	G\( Z \) = \Gbar \( 1 + \alpha \tanh\( \frac{Z - \Zbar}{L} \) \) 
\, ,
\label{eqn:gradation_O-centered}
\end{equation}
where $\Zbar = \refZbar - \refZ_O$.
The change of $\Zbar$ corresponds to the placement of $O$, center of the undeformed cavity, at different planes of $\refZ$ in the specimen-fixed frame.
When $\Zbar < 0$, $O$ is on the stiffer side of the specimen.
When $\Zbar > 0$, $O$ is on the more compliant side of the specimen. 
When $\Zbar = 0$, $O$ lies on the gradation midplane and we refer to such a cavity as gradation-centered.

\subsection{Cavity evolution and geometric characterization}
\label{subsec:formulation-cavity-geometry}

The cavity is inflated quasi-statically by a spatially uniform internal pressure.
Because the stiffness gradation and the applied loading both remain invariant under rotation about $\basis_3$, we restrict our analysis to axisymmetric deformations.
Accordingly, all fields are independent of $\Phi$ and material points are described simply by the reference radial coordinate $R$ and polar coordinate $\Theta$.

The inflated cavity has a current volume $V$. 
Due to the stiffness gradation in the surrounding material, the cavity is not expected to maintain sphericity.
However, as a measure of the cavity's volume change, we define a volume-equivalent cavity radius,
\begin{equation}
	a_V = \(\frac{3V}{4\pi}\)^{1/3} = \Lambda_{V} \, A
\, .	
\end{equation}
$\Lambda_{V}$ is the volume-equivalent stretch of the cavity surface, which we interpret as the level of inflation.

Another anticipated consequence of the stiffness gradation is that the centroid of the deformed cavity migrates away from $O$ to a spatial point $C$.
Denoting the current positions of $C$ and $O$ as $\bx_C$ and $\bx_O$, respectively, we define the migration amplitude to be the offset of $C$ from $O$,
\begin{equation}
	\delta = \(\bx_C - \bx_O\) \cdot \(-\basis_3\)
\, .
\label{eqn:delta_defn}
\end{equation}
Under uniform internal pressure, the more compliant side of the material offers less resistance to expansion than the stiffer side.
Therefore, we expect the centroid of the inflated cavity to migrate toward the compliant side of the material. 
\Cref{eqn:delta_defn} ensures that $\delta \geq 0$ when $\Gamma \geq 1$. 

To analyze the geometric distortion of the cavity wall, we consider a $C$-centered spherical frame in the current configuration, with radial coordinate $r_C$ and polar coordinate $\theta_C$, as illustrated in \Cref{fig:geometry-schematics}~(b).
The cavity wall is parametrized by $\theta_C$ and $r_C = a\(\theta_C\)$.
We quantify the cavity distortion by spectrally decomposing $a\(\theta_C\)$ as
\begin{equation}
	a\(\theta_C\) = \sum_{k = 0}^{\infty} a_k P_k\( \cos \theta_C \)
\, , \quad
    a_k = 
    \frac{2k + 1}{2} \int_{0}^{\pi} a\(\theta_C\) P_k\( \cos\theta_C \) \sin\theta_C \, \dd{\theta}_C
% Self reminder: in full form, negative sign is there because cos is smallest at \pi
\label{eqn:legendre-decomp}
\, ,
\end{equation}
where $P_k$ is the Legendre polynomial of degree $k$ and $a_k$ is the corresponding mode amplitude~\cite{arfken-weber-harris_2011_math-methods}.
The geometric distortion corresponding to mode degrees $k = $ 1 through 6 are illustrated in \Cref{fig:O-vs-C}~(a).
To leading order, the $k = 1$ mode corresponds to a translation along $\basis_3$, while the $k = 2$ mode represents a prolation-oblation along $\basis_3$, as seen most clearly in the darker-colored, $a_k/a_0 = 0.4$ cases in \Cref{fig:O-vs-C}~(a).
However, as $a_k/a_0$ increases in magnitude, the distortion modes deviate further from these simple geometric interpretations and higher-order shape effects become visually apparent, as in the lighter-colored, $a_k/a_0 = 0.8$ cases.

\subsection{Necessity of using $C$-centered coordinate frame for modal decomposition}
\label{subsec:leakage}

\begin{figure}[tb]
    \centering
    \includegraphics[width=0.75\linewidth]{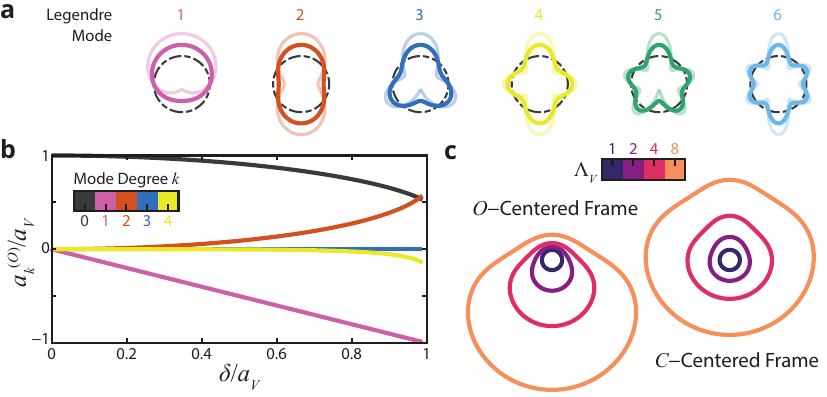}
    \caption{
    Modal decomposition of cavity geometry and the importance of using a $C$-centered spherical frame.
    (a) Illustration of individual Legendre distortion modes $k = $ 1 through 6, resulting in radial coordinate $r_C = a_0 + a_k P_k\(\cos\theta_C\)$.
    The darker curves correspond to relative amplitude $a_k/a_0 = 0.4$, while the lighter curves correspond to $a_k/a_0 = 0.8$.
    (b) Amplitudes of Legendre modes $k = $ 0 through 4 evaluated in $O$-centered frame for a purely translated sphere.
    While $a_1^{(O)} = -\delta$ is accurately recovered, the migration signal also leaks into other modes. 
    In contrast, a $C$-centered frame leads to $a_0 = 1$ and $a_k = 0$ for $k > 0$.
    (c) Deformed cavity geometry (from finite-element simulation of $\Gamma = 4, L^* = 1, \Zbar^* = 0$), shown in $O$- vs. $C$-centered frames, for varying levels of inflation.
    In the $C$-centered frame, the cavity's migration is decoupled from the nonspherical distortion.
    }
    \label{fig:O-vs-C}
\end{figure}

The choice of a $C$-centered frame is essential for decoupling the migration mode of the cavity from the nonspherical distortion modes.
To clearly illustrate this point, we consider here a spherical cavity of radius $a_V$ translated by $\delta = \eta \, \, a_V$ in the $-\basis_3$ direction, without additional nonspherical distortions.
If we were to work in a $O$-centered frame, we may define the corresponding current radial coordinate $r_O$ and polar coordinate $\theta_O$, as illustrated in \Cref{fig:geometry-schematics}~(b). 
The translated sphere satisfies
\begin{equation}
     \(r_O \cos\theta_O + \delta\)^2 + \(r_O \sin\theta_O \)^2  = {a_V}^2
\quad \Rightarrow \quad 
     r_O = {a_V} \( \sqrt{1 - \eta^2 \(\sin\theta_O\)^2} - \eta \cos\theta_O \)
     \, . 
\end{equation}
Performing Legendre decomposition of $r_O$, we obtain mode amplitudes $a_k^{(O)}$:
\begin{equation}
\begin{aligned}
     a_k^{(O)} &= \frac{2 k + 1}{2} \int_0^{\pi} r \(\theta_O\) P_k\( \cos\theta_O \) \sin\theta_O \, \dd{\theta_O}
\, , \\
     a_0^{(O)} &= {a_V}\(1 - \frac{\eta^2}{3} - \frac{\eta^4}{15} - \dots\)
\, , \qquad
     a_1^{(O)} = -\eta \, {a_V}
\, , \\
     a_2^{(O)} &= {a_V} \(\frac{\eta^2}{3} + \frac{2\eta^4}{21} + \dots\)
\, , \qquad
     a_3^{(O)} = 0
\, , \\
     a_4^{(O)} &= {a_V} \(-\frac{\eta^4}{35} - \frac{9\eta^6}{385} - \dots\)
\, , \qquad
     a_5^{(O)} = 0
\, , \\
    & \dots
\end{aligned}
\end{equation}
The resulting values of $a_k^{(O)}$ are plotted in \Cref{fig:O-vs-C}~(b) for $k = $ 0 through 4.
While $a_1$ matches $-\delta$ perfectly, we also recover non-zero amplitudes for other nonspherical distortion modes and a deviation of $a_0^{(O)}/a_V$ from unity, even though the cavity remains spherical.
In the $O$-centered Legendre decomposition, centroid migration contaminates Legendre modes other than $k = 1$.

When a $C$-centered frame is used, $r_C\(\theta_C\) = a_V$ at all points and we obtain, from \cref{eqn:legendre-decomp}, $a_0 = a_V$ and $a_k = 0$ for all Legendre modes with $k > 0$.
As further illustrated in \Cref{fig:O-vs-C}~(c), the $C$-centered Legendre modal decomposition properly decouples the centroid migration and nonspherical distortion of the cavity.

\subsection{Finite-element analysis}
\label{subsec:fe-intro}

At a prescribed volume-equivalent stretch $\Lambda_V$, the interaction between the inflated cavity and the surrounding graded material is fundamentally influenced by three dimensionless parameters: 
\begin{enumerate}[label=(\arabic*)]
    \item the stiffness contrast $\Gamma$;
    \item the normalized length scale of the gradation $L^* = L/A$;
    \item the normalized location of the gradation midplane $\Zbar^* = \Zbar/A$, which may be interpreted as the signed axial offset of the undeformed cavity with respect to the stiffness gradation.
\end{enumerate}

To explore the effects of these gradation parameters on the cavity geometry, we performed finite-element simulations using Abaqus/Standard~\cite{abaqus-standard}.
Details of the finite-element implementation are presented in \ref{app:fem}.
Briefly, the hyperelastic domain was modeled using a nearly incompressible penalty formulation, with a spatially uniform bulk modulus $K \gg \Gbar$, and user subroutines were developed to prescribe spatial gradation of the shear modulus $G\(Z\)$ according to \cref{eqn:gradation_O-centered}.
We choose a neo-Hookean form of strain energy density,
\begin{equation}
    \psi = \frac{G}{2} \(\Ibar_1 - 3\) + \frac{K}{2}\(J - 1\)^2
    \, ,
\label{eqn:nh-strain-energy}
\end{equation}
where $\Ibar_1$ is the first invariant of the modified Cauchy--Green deformation tensors and $J$ the Jacobian determinant of the deformation gradient~\cite{holzapfel_2000_nonlinear-solid-mechanics}.
The neo-Hookean model provides a simple constitutive setting in which the effect of the graded shear modulus can be isolated.
The cavity was inflated in a volume-controlled manner, following the procedure outlined in \citet{kolluri_2025_thin-IMR}.
At each loading increment, the nodal displacements of the cavity surface were used to compute the migration amplitude $\delta$ and the $C$-centered Legendre mode amplitudes $a_k$.

Simulations became more numerically challenging at larger stiffness contrast $\Gamma$, particularly when the normalized gradation length scale $L^*$ was on the order of unity.
As detailed in \ref{app:fem}, in some of these challenging cases, the default convergence criteria in Abaqus/Standard were not met after repeated automatic cutbacks once $\Lambda_V$ exceeded a case-dependent limit of robust convergence.
These cases are explicitly marked in the figures below, and results are reported only over the range of $\Lambda_V$ with robust convergence.
Because the convergence difficulties occurred only after appreciable cavity inflation and only for a small subset of the parameter space examined, they do not affect our main findings presented below.

% =====================
\section{Effects of gradation parameters on geometric modes}
\label{sec:fe-results}

\subsection{Centroid migration as the primary geometric response}
\label{subsec:fe_mode-evol}

\begin{figure}[tb]
    \centering
    \includegraphics[width=0.75\linewidth]{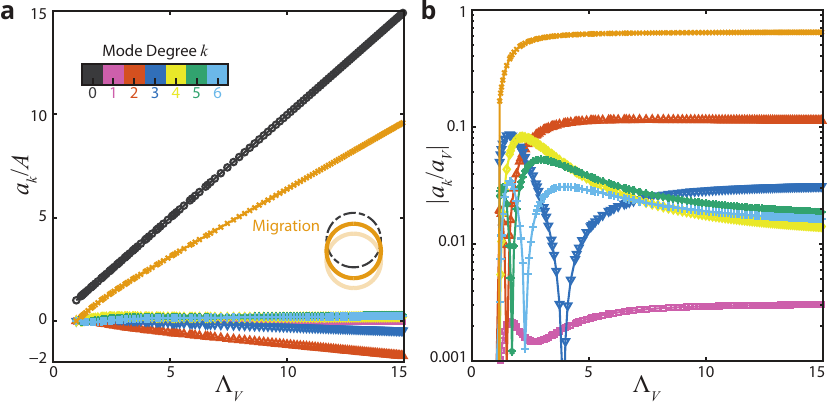}
    \caption{
    Finite-element simulation results for the representative case of $\Gamma = 4, L^* = 1, \Zbar^* = 0$:
    (a) mode amplitude normalized by reference radius $A$,
    (b) absolute value of mode amplitude normalized by current volume-equivalent radius $a_V$.
    For the purpose of comparison, we denote $a_k = \delta$ for the centroid migration.
    Beyond an initial small-inflation regime, the magnitude of centroid migration is generally larger than the magnitude of nonspherical distortion modes. 
    }
    \label{fig:fe-raw}
\end{figure}

To illustrate our observations from the finite-element simulations, we first consider a representative case defined by $\Gamma = 4, L^* = 1, \Zbar^* = 0$.
We subsequently show that the observed trends persist across the broader parameter range examined.

In \Cref{fig:fe-raw}, we present the evolution of the centroid migration and Legendre modes as $\Lambda_V$ is increased from $1$ to $15$ in the representative case.
The mode amplitudes are normalized by the reference radius $A$ in \Cref{fig:fe-raw}~(a) and by the current volume-equivalent radius $a_V$ in \Cref{fig:fe-raw}~(b).
In the $C$-centered frame, the $k=0$ mode amplitude $a_0$ closely matched the volume-equivalent radius $a_V$, while the $k=1$ mode amplitude $a_1$ remained negligible following the decoupling of centroid migration.
The $k=2$ mode was the most pronounced nonspherical distortion mode, perturbing the cavity geometry, to leading order, in the form of prolation-oblation.
However, beyond an initial range of small $\Lambda_V$ in which $\delta \ll A$ and $|a_k| \ll A$ for all $k>0$, the migration amplitude $\delta$ is clearly larger than the nonspherical distortion mode amplitudes.
We see in \Cref{fig:fe-raw}~(b) that, while $a_2$ increased above $a_V/10$ for sufficiently large $\Lambda_V$, it remained below $\delta/5$.
We did not observe $a_k > a_V/10$ for any $k > 2$ modes in the range of volume-equivalent stretch $\Lambda_V$ and gradation parameters investigated.
The $O$-centered and $C$-centered geometry for this representative case are shown in \Cref{fig:O-vs-C}~(c) for various inflation levels.
While the inflated cavity no longer remained spherical, the amplitudes of the nonspherical distortion modes in the $C$-centered frame are apparently small compared to that of the centroid migration.

Some additional representative cases are presented in \ref{app:more-fe-results}.
Across the broad range of $\Gamma$, $L^*$, and $\Zbar^*$ examined, the finite-inflation magnitude of centroid migration was generally several times larger than that of any nonspherical distortion mode.
This suggests that the cavity's centroid migration is the strongest geometric signature of the mechanical gradation.

To further examine the relation between centroid inflation and migration, we introduce the migration ratio
\begin{equation}
    \eta = \delta/a_V
\end{equation}
as a measure of the proportion of migration versus inflation experienced by the cavity.
For the representative case shown in \Cref{fig:fe-raw}, $\eta$ increased monotonically with increasing $\Lambda_V$ and then reached a plateau for $\Lambda_V \gtrsim 5$.
For some combinations of gradation parameters (e.g., \Cref{fig:eta-gamma}~(b)), $\eta$ decreased slightly with increasing $\Lambda_V$ prior to the plateau regime, even though the dimensional migration amplitude $\delta$ continued to increase monotonically.
The sensitivity of $\eta$ to the gradation parameters, $\Gamma$, $L^*$, and $\Zbar^*$, is the main focus of our analysis below.

\subsection{Effect of $\Gamma$ on the migration of gradation-centered cavity}
\label{subsec:fe_eta-gamma}

\begin{figure}[tb]
    \centering
    \includegraphics[width=0.75\linewidth]{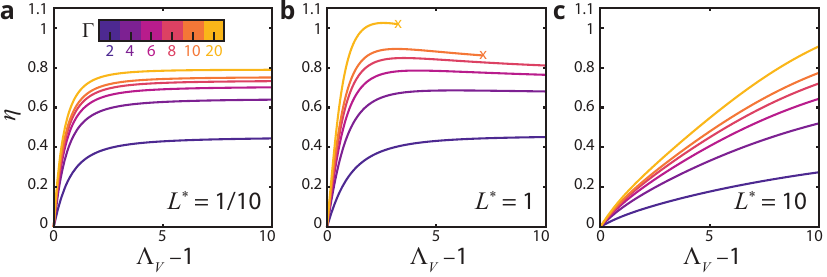}
    \caption{
    Effect of stiffness contrast $\Gamma$ on the evolution of migration ratio $\eta$ when $\Zbar^* = 0$ and (a) $L^* = 1/10$, (b) $L^* = 1$, (c) $L^* = 10$.
    Across steep, moderate, and gradual stiffness gradients, $\Gamma$ monotonically increased $\eta$ at a prescribed $\Lambda_V$.
    For the $\Gamma = 10$ and $\Gamma = 20$ cases with $L^* = 1$, the finite-element simulations failed to meet convergence criteria for $\Lambda_V > 8.1$ and $\Lambda_V > 4.2$, respectively.
    }
    \label{fig:eta-gamma}
\end{figure}

We first draw our attention to cases with $\Zbar^* = 0$, in which the center of the undeformed cavity lies on the gradation midplane.
The effect of stiffness contrast $\Gamma$ on the evolution of $\eta$ is illustrated in \Cref{fig:eta-gamma}.
Consistent across stiffness gradation profiles with $L^* = 1/10$, $1$, and $10$, we observe that an increased stiffness contrast $\Gamma$ resulted in a larger $\eta$ at a fixed $\Lambda_V$.
Although the growth of $\eta$ is not necessarily monotonic with increasing $\Lambda_V$, as seen in \Cref{fig:eta-gamma}~(b) when $L^* = 1$ and $\Gamma > 4$, the increase of $\Gamma$ still corresponded to a monotonic increase of $\eta$ in this range of moderate $\Lambda_V$.

\subsection{Effect of $L^*$ on the migration of gradation-centered cavity}
\label{subsec:fe_eta-length}

\begin{figure}[tb]
    \centering
    \includegraphics[width=0.75\linewidth]{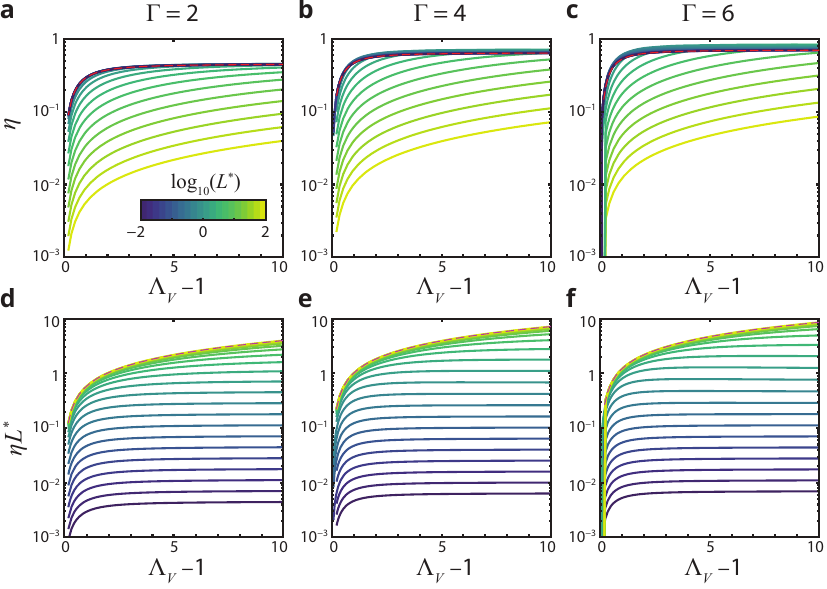}
    \caption{
    Effect of normalized gradation length scale $L^*$ for $\Zbar^* = 0$.
    The increase of $L^*$ at fixed $\Gamma$ generally leads to a decrease of $\eta$ at a fixed $\Gamma$ and a small to moderate $\Lambda_V$, prior to the plateau regime.
    The evolution of $\eta$ for (a) $\Gamma = 2$, (b) $\Gamma = 4$, and (c) $\Gamma = 6$ reveals an asymptotic limit (red dashed curve) for very steep stiffness gradients, with $L^* \ll 1$.
    The evolution of $\eta \, L^*$ for (d) $\Gamma = 2$, (e) $\Gamma = 4$ and (f) $\Gamma = 6$ reveals an asymptotic limit (magenta dashed curve) for very gradual stiffness gradients, with $L^* \gg 1$.
    }
    \label{fig:eta-Lstar}
\end{figure}

We next consider the effect of $L^*$ on the evolution of $\eta$ when $\Zbar^* = 0$, as illustrated in \Cref{fig:eta-Lstar} for the representative cases of $\Gamma = 2$, $4$, and $6$.

As $L^*$ decreases, the stiffness transition becomes steeper relative to the cavity size.
Within a low-$\Lambda_V$ inflation regime, the decrease of $L^*$ corresponded to a monotonic increase of $\eta$.
However, when the inflation level reached a moderate range (e.g., at $\Lambda_V \approx 2$ for the $\Gamma = 6$ case shown in \Cref{fig:eta-Lstar}~(c)), a very steep stiffness gradient with $L^* = 1/100$ resulted in a plateau of $\eta$, while a moderate stiffness gradient with $L^* = 1$ led to $\eta$ that continued to increase with larger $\Lambda_V$ and surpassed that of the $L^* = 1/100$ case.

When $L^* \ll 1$, the evolution of $\eta$ exhibited an asymptotic behavior, as shown in \Cref{fig:eta-Lstar}~(a)--(c), with the $L^* \leq 1/10$ curves becoming indiscernible from the red, dashed curve.
This suggests that, in the steep-gradient limit, the evolution of $\eta$ becomes less sensitive to $L^*$ and reflects the effect of $\Gamma$ more clearly.

Another asymptotic behavior is observed when $L^* \gg 1$.
As shown in \Cref{fig:eta-Lstar}~(d)--(f), the quantity $\eta \, L^*$ generally increased when $L^*$ increased. 
However, an increase of $L^*$ beyond $10$ resulted in diminishing changes to the evolution of $\eta \, L^*$ within the range of $\Lambda_V$ examined, with the $L^* \geq 10$ curves becoming indiscernible from the magenta, dashed curve.
In other words, in the gradual-gradient limit, the migration ratio $\eta$ is approximately inversely proportional to $L^*$.

Considering together the results presented in \Cref{fig:eta-gamma} and \Cref{fig:eta-Lstar}, we further observed that the large-inflation plateau value of $\eta$ remained consistent across different $L^*$ when $\Gamma$ was fixed. 
The plateau value increased monotonically with $\Gamma$, but, as seen in \Cref{fig:eta-gamma}~(a), the increase of the plateau value saturated as $\Gamma$ became very large and $\alpha = \(\Gamma - 1\)/\(\Gamma + 1\)$ approached the limit value of $1$.

\subsection{Leading-order symmetry of $\eta$ about $\Zbar^*$}
\label{subsec:fe_eta-symm}

\begin{figure}[tb]
    \centering
    \includegraphics[width=0.75\linewidth]{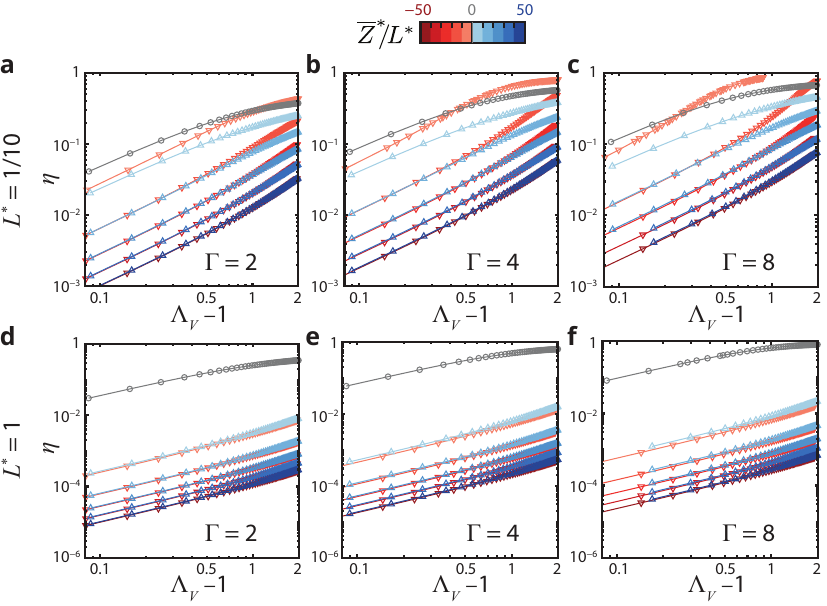}
    \caption{
    Effect of $\Zbar^*$ on the evolution of $\eta$:
    steep stiffness gradient with $L^* = 1/10$ and stiffness contrast (a) $\Gamma = 2$, (b) $\Gamma = 4$, (c) $\Gamma = 8$;
    moderate stiffness gradient with $L^* = 1$ and stiffness contrast (d) $\Gamma = 2$, (e) $\Gamma = 4$, (f) $\Gamma = 8$.
    The evolution of $\eta$ during the initial, low-$\Lambda_V$ regime was approximately symmetric about $\Zbar^* = 0$, with a more gradual increase of $\eta$ (i.e., smaller $\partial{\eta}/\partial{\Lambda_V}$) with larger $|\Zbar^*|$.
    The departure from the leading-order symmetry was more pronounced when $\Lambda_V$ increased and when $|\Zbar^*|$ decreased.
    }
    \label{fig:eta-Zbar}
\end{figure}

We now examine the role of $\Zbar^*$ in the evolution of $\eta$ when $\Gamma$ and $L^*$ are held constant. 
We performed finite-element simulations with $\Zbar^*/L^*$ varying from $-50$ to $50$, corresponding to a relocation of $O$ from the stiffer half of the graded solid ($G > \Gbar$) to the more compliant half ($G < \Gbar$).
The results are shown in \Cref{fig:eta-Zbar} for representative combinations of $\Gamma$ and $L^*$.

For the initial growth stage corresponding to small values of $\Lambda_V$, we observe that the migration ratio $\eta$ was largest when $\Zbar^* = 0$.
As the center of the undeformed cavity was placed farther away from the gradation midplane, with increasing $|\Zbar^*|$, the migration amplitude diminished monotonically in magnitude at a prescribed $\Lambda_V$.
Remarkably, the magnitude of $\eta$ in this regime was not only maximized at around $\Zbar^* = 0$ but also exhibited an apparent symmetry.
For example, $\Zbar^*/L^* = -20$ and $\Zbar^*/L^* = 20$ resulted in nearly identical $\eta\(\Lambda_V\)$ curves in the small-inflation regime.
The apparent leading-order symmetry of $\eta$ with respect to $\Zbar^*$ persisted when $L^*$ was further increased or decreased beyond the range shown in \Cref{fig:eta-Zbar}.
This apparent symmetry provides a robust target for us to qualitatively identify the gradation midplane, $\refZ = \refZbar^*$, within the specimen.

When $\Lambda_V$ increased, the symmetry about $\Zbar^* = 0$ was broken more appreciably, with negative-valued $\Zbar^*$ generally leading to a larger $\eta$ when $|\Zbar^*|$, $\Gamma$, and $L^*$ were held constant.
This can be seen, for example, when comparing the $\Zbar^*/L^* = -10$ and $\Zbar^*/L^* = 10$ curves in \Cref{fig:eta-Zbar}~(a)--(c).

% =====================
\section{First-order Rayleigh-Ritz model for inflated-and-translated sphere}
\label{sec:first-order}

\begin{figure}[tb]
    \centering
    \includegraphics[width=0.75\linewidth]{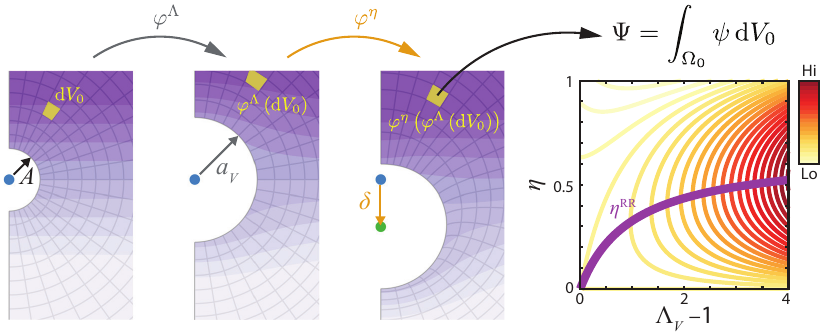}
    \caption{
    Outline of the Rayleigh--Ritz reduced-order model.
    We approximate the deformation mapping as the composition of cavity inflation $\defmap^{\Lambda}$ and cavity migration $\defmap^{\eta}$.
    The inflation map results in a cavity with radius $a_V = \Lambda_V \, A$, and the migration map then translates the cavity by a magnitude of $\delta = \eta \, a_V$ in the $-\basis_3$ direction.
    By integrating the graded hyperelastic strain energy density over the reference solid domain, we calculate the total strain energy $\Psi$ corresponding to the assumed deformation mapping and the stiffness gradation profile.
    For a prescribed volume-equivalent stretch $\Lambda_V$, the Rayleigh--Ritz solution of migration ratio $\eta^{\rm RR}$ minimizes $\Psi$. 
    }
    \label{fig:rr-scheme}
\end{figure}

The modal decomposition of the inflated cavity geometry revealed that, beyond an initial low-$\Lambda_V$ inflation regime with insignificant asymmetry, the most prominent form of asymmetry is generally the centroid migration.
This observation is consistent with a local Taylor expansion of the stiffness gradation \cref{eqn:gradation_O-centered} about the cavity center: to leading order, $G\(Z\) - G\(0\)$ is proportional to $Z = R \, P_1\(\cos\Theta\)$, a degree-one perturbation that couples naturally to centroid migration.
Furthermore, the migration ratio $\eta$ exhibited systematic trends that distinctively reflected the individual contributions of the gradation parameters $\Gamma$, $L^*$, and $\Zbar^*$.

In this section, we introduce a reduced-order model to estimate the dependence of the migration ratio $\eta$ on the volume-equivalent stretch $\Lambda_V$ and the gradation parameters.
The establishment of such an analytical, quantitative framework presents a route to isolate the contributions of each gradation parameter and, accordingly, infer mechanical properties from cavity geometry.

Because our finite-element simulations revealed comparatively weak nonspherical distortion during cavity inflation, we approximate the deformed cavity to be an inflated-and-translated sphere.
In other words, the cavity surface in the current configuration is represented by a sphere inflated from a undeformed radius of $A$ to a radius of $a_V = \Lambda_V A$ and then translated by a magnitude of $\delta = \eta \, a_V$ along the $-\basis_3$ direction, as illustrated in \Cref{fig:rr-scheme}.
This is equivalent to decomposing the deformation into a sequence of two mappings, the cavity inflation $\defmap^{\Lambda}$ and the cavity migration $\defmap^{\eta}$.
While this construction is not intended to perfectly represent the full deformation field, it provides a simplified framework to decouple the inflation and migration modes and rationalize the results presented in \Cref{sec:fe-results}. 

We demonstrate below that, within the simplified context of homogeneous linear elasticity, the displacement fields resulting from both the pure inflation and pure translation of a spherical cavity are analytically tractable with standard techniques~\cite{sadd_2025_elasticity}.
Using these two sets of kinematically-admissible displacement fields as bases, we then construct a family of trial displacement fields approximating the response of a graded material containing an inflated cavity that migrates.
Following a Rayleigh--Ritz process, we solve for $\eta$ that globally minimizes the strain energy for prescribed $\Lambda_V$ and gradation parameters.

\subsection{Auxiliary linear elastic solutions}

In the linear elastic solutions presented below and in \ref{app:translate-sphere-soln}, the spherical coordinate system remains centered at the initial cavity center $O$, so that cavity translation is represented explicitly through the displacement field.
For brevity, we will directly denote the current radial and polar coordinates as $r = r_O$ and $\theta = \theta_O$.

We consider an unbounded body of incompressible linear elastic material, containing a sphere of radius $A$.
When the sphere is inflated to a radius of $a_V = \Lambda_V \, A$, a material point with reference radial coordinate $R$ is displaced to a current radial coordinate $r$, while the current polar and azimuthal coordinates satisfy $\theta = \Theta$ and $\phi = \Phi$, respectively.
In this case, the small strain tensor $\beps$ has the non-zero components
\begin{equation}
   \epsilon_{RR} = \fracp{r}{R} - 1
\, , \quad
    \epsilon_{\Theta\Theta} = \epsilon_{\Phi\Phi} = \frac{r}{R} - 1 
\, .
\end{equation}
Incompressibility requires that $\tr{\beps} = 0$, which then leads us to
\begin{equation}
   u_R = \frac{A^3\(\Lambda_V - 1\)}{R^2}
\, ,
\label{eqn:infl_u}
\end{equation}
while $u_{\Theta} = u_{\Phi} = 0$.

Next, we consider the case where the sphere is translated by $\delta$ along the $-\basis_3$ direction in the unbounded linear elastic body.
The surface of the sphere is assumed to be shear-free (i.e., ``slippery''), such that traction is purely radial.
The solution of this boundary value problem is reviewed in \ref{app:translate-sphere-soln}. 
We find that the corresponding components of the displacement field are
 \begin{equation}
    u_R = -\frac{\delta \, A}{R} \cos\Theta
\, , \quad
    u_{\Theta} = \frac{\delta \, A}{2 R} \sin\Theta
\, , \quad
    u_{\Phi} = 0
\, .
\label{eqn:trans_u}
\end{equation}

% ===============
\subsection{Rayleigh--Ritz minimization of strain energy}
\label{subsec:ro-first}

We follow a Rayleigh--Ritz process~\cite{zienkiewicz-taylor-govindjee_2024_fem} to optimize $\eta = \delta/a_V$ for a given value of $\Lambda_V$ in a graded material with an undeformed cavity satisfying $\Zbar^* = 0$.
We linearly superpose the kinematically-admissible displacement fields in \cref{eqn:infl_u,eqn:trans_u} as basis modes and seek $\eta$ that globally minimizes the strain energy in the unbounded body when $\Lambda_V$ is prescribed.

From the infinitesimal strain components corresponding to the sum of displacement fields \cref{eqn:infl_u,eqn:trans_u}, we find the local strain energy density to be
\begin{equation}
   \psi\(R,\Theta\) % = G\(\epsilon_{ij} \epsilon_{ij}\) 
= \frac{3\,G\(R,\Theta\)\, A^2}{2\,R^6} \(R \, \delta \cos\Theta - 2 \, A^2 \(\Lambda_V - 1\)\)^2
\, .
\end{equation}
For the unbounded body of graded material occupying the domain $\Omega_0$ in the reference configuration, we obtain the total strain energy by integrating $\psi$ over $\Omega_0$, % with reference volume element $\dd{V}_0$,
\begin{equation}
    \Psi = \int_{\Omega_0} \psi \, \dd{V}_0 
= 2\pi 
\int_{A}^{\infty}\int_{0}^{\pi} \psi\(R,\Theta\) 
R^2 \sin\Theta \, 
\dd{\Theta} \, \dd{R}
\, .
\end{equation}

Since $\psi$ is quadratic with respect to $\delta$, the minimization of $\Psi$ with respect to $\delta$ is equivalent to seeking $\partial\Psi/\partial{\delta} = 0$. 
For this purpose, it is convenient to first find the local derivative of strain energy density with respect to $\delta$,
\begin{equation}
     \frac{\partial\psi}{\partial\delta} = 
3 \, G\(R,\Theta\) \, A^2 \, R^{-5} \cos\Theta\(R \, \delta \cos\Theta - 2A^2\(\Lambda_V - 1\)\)
\, .
\label{eqn:energy-density-min}
\end{equation}
To quantitatively delineate the roles of $L^*$ and $\Gamma$ in the evolution of $\eta$, we first focus on the case of $\Zbar^* = 0$.
To facilitate the calculation, we introduce the radial point mapping $\xi = 1 - 2A/R$ and denote $\cos\Theta = c$ and $\sin\Theta = \sqrt{1-c^2} = -\dd{c}/\dd\Theta$, 
such that $G = \overline{G}\(1 + \alpha \tanh\(c\,R/L\)\)$ for $\Zbar^* = 0$. 
We then obtain
\begin{equation}
\begin{aligned}
    \fracp{\Psi}{\delta} &=  \(4\pi \, \Gbar \, A^2\)\( \eta \, \Lambda_V - \alpha\(\Lambda_V - 1\) H\( L^* \) \)
\, .
\label{eqn:energy-min-condition}
\end{aligned}
\end{equation}
where 
\begin{equation}
    H\( L^* \) = \int_{-1}^{1} \int_{-1}^{1} 
      \frac{3 \, c}{4}\(1 - \xi\)  \tanh\(\frac{2 \, c}{L^*\(1 - \xi\)}\)
    \dd{c} \, \dd{\xi}
\end{equation}
is a quantity that may be accurately evaluated with Gauss--Legendre quadrature.
In our analysis, we used 256-point quadrature for both $\xi$ and $c$.

From \cref{eqn:energy-min-condition}, we obtain the first-order Rayleigh--Ritz solution of cavity migration that minimizes the strain energy $\Psi$,
\begin{equation}
    \eta^{\rm RR1} = \alpha \(1 - \Lambda_V^{-1}\) H\( L^* \)
\, .
\label{eqn:ro-1_eta}
\end{equation}
When $\Lambda_V$ increases, $\eta^{\rm RR1} $ increases monotonically.
When the stiffness contrast $\Gamma$ increases, we have a monotonic increase of $\alpha$ leading to a monotonic increase of $\eta^{\rm RR1} $.
The $\Lambda_V$- and $\Gamma$-dependent trends qualitatively match our finite-element simulation results presented in \Cref{sec:fe-results}.

As the stiffness gradation profile approaches the steep- and gradual-gradient limits, respectively, we have 
\begin{equation}
\begin{aligned}
    \lim_{L^* \to 0} H &= \frac{3}{2}
    \quad \Rightarrow &
    \lim_{L^* \to 0} \eta^{\rm RR1}  &= \frac{3 \alpha}{2}\(1 - \frac{1}{\Lambda_V}\)
\, , \\
    \lim_{L^* \to \infty} H &= \frac{2}{L^*}
    \quad \Rightarrow &
    \lim_{L^* \to \infty} \eta^{\rm RR1}  &= \frac{2 \alpha}{L^*}\(1 - \frac{1}{\Lambda_V}\)
\, .
\label{eqn:ro1-limits}
\end{aligned}
\end{equation}
We recover the independence of $\eta^{\rm RR1}$ to $L^*$ in the steep-gradient limit and the inverse correlation of $\eta^{\rm RR1}$ to $L^*$ in the gradual-gradient limit, again matching the trends observed in the finite-element simulations.

% ===============
\subsection{Breakdown of the first-order model at finite inflation}

\begin{figure}[tb]
    \centering
    \includegraphics[width=0.75\linewidth]{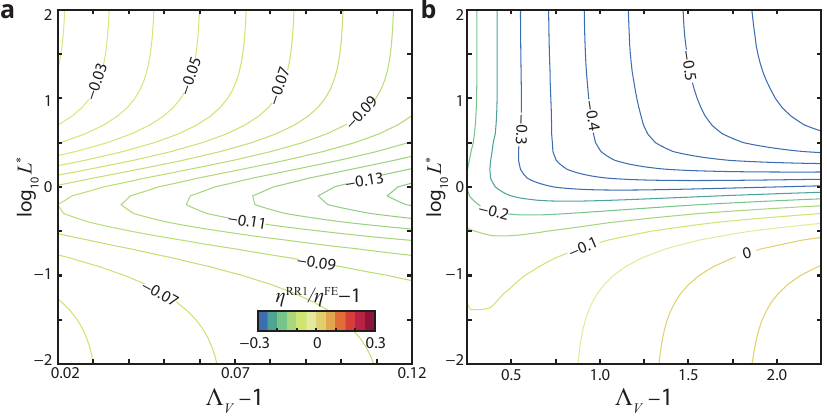}
    \caption{
    Signed error of the first-order Rayleigh--Ritz solution $\eta^{\rm RR1}$ relative to finite-element solution $\eta^{\rm FE}$, for the representative case of $\Gamma = 4$, $\Zbar^* = 0$.
    (a) For small inflation, the first-order Rayleigh--Ritz solution slightly underpredicts the cavity migration, with a relative error magnitude within $13\%$ for $\Lambda < 1.1$.
    (b) For finite inflation, the accuracy of $\eta^{\rm RR1}$ deteriorates, notably with relative error magnitude exceeding $50\%$ at $\Lambda_V = 2.5$ for $L^* = 1/100$.
    This points to an omitted finite-deformation effect.
    }
    \label{fig:ro-1_eta-evol}
\end{figure}

To assess the capability of the first-order Rayleigh--Ritz model to quantify the evolution of $\eta$ when $\Zbar^* = 0$, we examine the relative error of the first-order Rayleigh--Ritz solution $\eta^{\rm RR1}$ with respect to the finite-element simulations results, which we denote as $\eta^{\rm FE}$.

In \Cref{fig:ro-1_eta-evol}, we present the signed relative error, $\eta^{\rm RR1}/\eta^{\rm FE} - 1$, for the representative case of $\Gamma = 4$.
When the volume-equivalent stretch $\Lambda_V$ is near the undeformed limit of $1$, as shown in \Cref{fig:ro-1_eta-evol}~(a), $\eta^{\rm RR1}$ generally underpredicts the migration ratio, though with a small error magnitude relative to $\eta^{\rm FE}$.
The agreement is particularly good when the stiffness gradient is very gradual, with $L^* > 10$, or when the stiffness gradient is very steep, with $L^* < 1/10$.
For these cases, the relative error magnitude is generally within $8\%$ for $\Lambda_V < 1.1$.

As $\Lambda_V$ increases, the disagreement between the two solutions increases drastically for gradual stiffness gradient with large $L^*$.
As shown in \Cref{fig:ro-1_eta-evol}~(b), when $\Lambda_V = 2.5$, $\eta^{\rm RR}$ underpredicts $\eta^{\rm FE}$ by over $50\%$ for $L^* = 100$.
Interestingly, for smaller $L^*$, the difference between $\eta^{\rm RR1}$ and $\eta^{\rm FE}$ does not remain monotonic in a moderate inflation regime.
In the case of $L^* = 1/100$, we see in \Cref{fig:ro-1_eta-evol}~(b) that $\eta^{\rm RR1}$ accurately matches $\eta^{\rm FE}$ near $\Lambda_V = 2.3$.
When $\Lambda_V$ further increases, however, the magnitude of relative error increases again.

Across different levels of stiffness contrast $\Gamma$ examined, the predictability of the first-order Rayleigh--Ritz solution consistently breaks down with larger $\Lambda_V$.
The first-order approximation based on linear elastic displacement fields become less effective when the inflation level is large and deformation is no longer infinitesimal.

% ===============

\subsection{Leading-order symmetry about $\Zbar^* = 0$}
\label{subsec:ro1_eta-symm}

\begin{figure}[tb]
    \centering
    \includegraphics[width=0.75\linewidth]{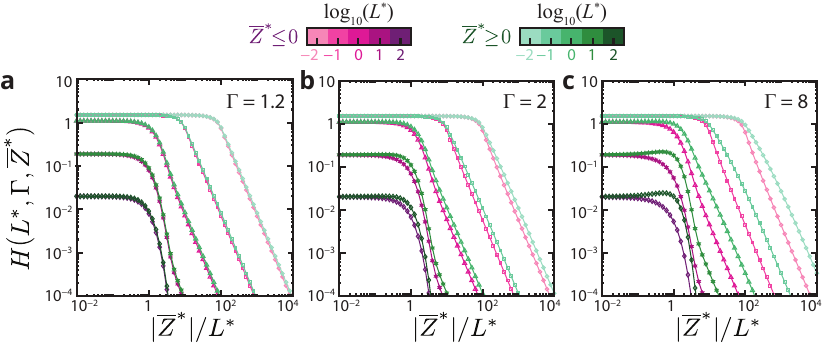}
    \caption{
    Numerical values of $H$ in the generalized first-order Rayleigh--Ritz model, calculated via \Cref{eqn:KQ_linear-general}, for the representative cases of (a) $\Gamma = 1.2$, (b) $\Gamma = 2$, and (c) $\Gamma = 8$.
    For a prescribed pair of stiffness contrast $\Gamma$ and normalized gradation length scale $L^*$, $H$ exhibits leading-order symmetry about $\Zbar^* = 0$.
    Accordingly, the first-order model predicts a leading-order symmetry of the migration ratio $\eta$ about $\Zbar^* = 0$.
    The departure from the leading-order symmetry becomes more pronounced when $\Gamma$ and $|\Zbar^*|$ are increased.
    }
    \label{fig:ro-1_eta-symm}
\end{figure}

In \Cref{subsec:fe_eta-symm}, we identified a qualitative dependence of the small-inflation response on the normalized location of the gradation midplane, $\Zbar^*$. 
In finite element simulations with fixed stiffness contrast $\Gamma$ and normalized gradation length scale $L^*$, the initial growth rate of $\eta$ with volume-equivalent stretch $\Lambda_V$ was typically greatest when the undeformed cavity was gradation-centered, with $\Zbar^* = 0$.
The initial growth rate was also approximately symmetric about $\Zbar^* = 0$. 
Here, we use the first-order Rayleigh--Ritz model to isolate the leading mechanism underlying this apparent symmetry.

We may generalize the procedure leading to \cref{eqn:ro-1_eta} by considering an offset of the cavity relative to the gradation midplane, with $\Zbar^* \neq 0$.
Substituting \cref{eqn:gradation_O-centered} in \cref{eqn:energy-density-min}, we obtain
\begin{equation}
     \frac{\partial\psi}{\partial\delta} = 
3 \, \Gbar \(1 + \alpha \, \tanh\(\frac{\cos\Theta \, R/A - \Zbar^*}{L^*}\) \) \, A^2 \, R^{-5} \cos\Theta\(R \, \delta \cos\Theta - 2A^2\(\Lambda_V - 1\)\)
\, .
\end{equation}
Then, integrating over the reference solid domain, we arrive at a generalization of the result shown in \cref{eqn:energy-min-condition},
\begin{equation}
    \fracp{\Psi}{\delta} = 3\pi \, \overline{G} A^2 \( \eta \, \altK \, \Lambda_V - \alpha \, \altQ \(\Lambda_V - 1\) \)
\, ,
\end{equation}
where
\begin{equation}
\begin{aligned}
\altQ &= \int_{-1}^{1} \int_{-1}^{1} 
        \(1 - \xi\) c  \tanh\( \(2c/\(1-\xi\) - \overline{Z}^*\)/L^* \)
\dd{c} \, \dd{\xi}
\, , 
\\
  \altK &= \(
\int_{-1}^{1} \int_{-1}^{1} 
     c^2 \,
\dd{c} \, \dd{\xi}
+
\alpha 
\int_{-1}^{1} \int_{-1}^{1} 
       {   c^2 \tanh\(\(2c/\(1-\xi\) - \overline{Z}^*\)/L^*\)}
\dd{c} \, \dd{\xi}
\) 
\, .
\label{eqn:KQ_linear-general}
\end{aligned}
\end{equation}
We then find the migration ratio to be $\eta^{\rm RR1} = \alpha \, H \(1 - \Lambda_V^{-1}\)$, where $H = \altQ/\altK$.

For the special case of $\Zbar^* = 0$, $\altK = 4/3$ and we recover the earlier result shown in \cref{eqn:ro-1_eta}, with $H$ dependent only on $L^*$.
Otherwise, $H$ depends on all three of the gradation parameters, $\Gamma$, $L^*$, and $\Zbar^*$.
This is illustrated in \Cref{fig:ro-1_eta-symm} for the cases of $\Gamma = 1.1$, $2$, and $8$, with varying $L^*$ and $\Zbar^*$.

When the stiffness contrast is low, as in the case of $\Gamma = 1.1$, the dependence of $H$ on the perturbation of $\Zbar^*$ from $0$ is relatively insensitive to the sign of $\Zbar^*$.
We see in \Cref{fig:ro-1_eta-symm}~(a) that relocating the cavity centroid by the same distance along $\basis_3$ and $-\basis_3$ results in a similar change to $H$.
The increase of $|\Zbar^*|$ generally leads to a decrease of $H$, which corresponds to a decrease of $\eta^{\rm RR1}$.
The leading-order symmetry of $\eta^{\rm RR1}$ with respect to $\Zbar^* = 0$ matches the finite-element simulation results in \Cref{subsec:fe_eta-symm}.

As $\Gamma$ increases, the symmetry of $\eta^{\rm RR1}$ is broken more noticeably, as in the cases of $\Gamma = 2$ and $\Gamma = 8$ shown in \Cref{fig:ro-1_eta-symm}~(b) and (c).
Contrary to the finite-element simulation results, the Rayleigh--Ritz model predicts a greater migration for a positive-valued $\Zbar^*$ than for a negative-valued $\Zbar^*$ when $|\Zbar^*|$, $L^*$, $\Gamma$, and $\Lambda_V$ are held constant.
This discrepancy between the Rayleigh--Ritz and finite-element results may reflect additional, higher-order dependence of $\eta$ on $\alpha$ that is not fully captured by the present Rayleigh--Ritz model.

% =====================
\section{Second-order Rayleigh--Ritz model for inflated-and-translated sphere}
\label{sec:second-order}

The first-order Rayleigh--Ritz model captures the small-inflation migration response and rationalizes its approximate symmetry about the gradation midplane. 
However, reducing the disagreement between the finite-element and Rayleigh--Ritz  predictions at higher inflation levels requires a treatment of the geometric nonlinearities associated with finite deformation.
To this end, we develop a second-order Rayleigh--Ritz model that incorporates the leading nonlinear kinematic corrections.

\subsection{Second-order approximation of finite deformation}

We consider a decomposition of the deformation mapping as depicted in \Cref{fig:rr-scheme}. 
The cavity inflation $\defmap^{\Lambda}$ maps the reference spherical coordinates $\{R, \Theta, \Phi \}$ to an intermediate set of $O$-centered coordinates $\{\hat{R}, \hat{\Theta}, \hat{\Phi} \}$. 
The cavity migration $\defmap^{\eta}$ then maps $\{\hat{R}, \hat{\Theta}, \hat{\Phi} \}$ to the current set of $O$-centered coordinates $\{r, {\theta}, {\phi} \}$.
Accordingly, the deformation gradient is written in a multiplicative decomposition form as 
\begin{equation}
    \defgrad = {\defgrad}^{\eta} \, {\defgrad}^{\Lambda}
    \, ,
    \label{eqn:F_coupled}
\end{equation}
allowing us to define a family of trial deformation mappings described by two parameters, the volume-equivalent stretch $\Lambda_V$ and the migration ratio $\eta$. 

The finite inflation of a spherical cavity in a neo-Hookean solid has a well-known solution~\cite{zhu-etal_2011_snap-through, estrada-etal_2018_IMR}, with
\begin{equation}
    u^{\Lambda}_{R} = \(R^3 + A^3\(\Lambda_V^3 - 1\)\)^{1/3} - R
    \, .
\label{eqn:u_finite-inflation}
\end{equation}
We see here that the linear elastic solution in \cref{eqn:infl_u} describes the leading-order effect of $\Lambda_V$ in \cref{eqn:u_finite-inflation}.
From $u^{\Lambda}_{R}$, we find the non-zero components of the inflation-related distortion ${\defgrad}^{\Lambda}$,
\begin{equation}
    F^{\Lambda}_{\hat{R} R} = \lambda^{-2}
    \, , \quad
    F^{\Lambda}_{\hat{\Theta} \Theta} = F^{\Lambda}_{\hat{\Phi} \Phi} = \lambda
    \, , \quad
    \lambda = \(\(\Lambda^3_V - 1\) \(A/R\)^3 + 1\)^{1/3}
    \, .
    \label{eqn:F_infl}
\end{equation} 

To the best of our knowledge, an exact solution for the translated sphere problem is not available for finite deformation.
However, we can follow a classical, series-expansion approach~\cite{rivlin_1953_second-order,green-adkins_1970_large-elastic-deformations} and develop a second-order approximation of the displacement field.
Specifically, for a sphere with current radius $a_{V}$, translated by a magnitude $\delta$, we approximate the migration-induced displacement field as
\begin{equation}
    \bu^{\eta} = \eta \, \bv + \eta^2 \, \bw 
    \, ,
    \label{eqn:series_u}
\end{equation}
where the migration ratio $\eta = \delta/a_V$ may be treated as a perturbation parameter.
The normalized first-order displacement, $\bv$, is available from our earlier, linear-elastic solution, while the normalized second-order displacement, $\bw$, is found in \ref{app:translate-sphere-second-order}.
Accordingly, the translation-related distortion ${\defgrad}^{\eta}$ has non-zero components, 
\begin{equation}
\begin{aligned}
     F^{\eta}_{r \hat{R}} &= 1 + \frac{\eta}{R^{*2}} \, \cos\Theta + \frac{\eta^2}{40 \, R^{*5}}\( 10 \, R^* + \(16 - 5 \, R^* - 8 \, R^{* 2}\)\(3 \(\cos\Theta\)^2 - 1\) \) 
\, ,
\\
     F^{\eta}_{r \hat{\Theta}} &= \frac{\eta}{2 \, R^{*2}}\sin\Theta + \frac{\eta^2}{40 \, R^{*5}}\(32 - 25 R^* - 24 R^{*2}\)\cos\Theta \, \sin\Theta
\, ,
\\
     F^{\eta}_{\theta \hat{R}} &= -\frac{\eta}{2 \, R^{*2}}\sin\Theta + \frac{\eta^2}{40 \, R^{*5}}\(32 - 45 R^*\)\cos\Theta \, \sin\Theta
\, ,
\\
     F^{\eta}_{\theta \hat{\Theta}} &= 1 - \frac{\eta}{2 \, R^{*2}}\cos\Theta + \frac{\eta^2}{240 \, R^{*5}}\( \(16 - 50 R^*\) - \(56 - 70 R^* - 24 R^{*2}\)\(3 \(\cos\Theta\)^2 - 1\) \)
\, ,
\\
     F^{\eta}_{\phi \hat{\Phi}} &= 1 - \frac{\eta}{2 \, R^{*2}}\cos\Theta + \frac{\eta^2}{120 \, R^{*5}}\( \(-8 + 5 R^*\) - \(20 - 20 R^* - 12 R^{*2}\)\(3 \(\cos\Theta\)^2 - 1\) \)
\, ,
\end{aligned}
\label{eqn:F_trans}
\end{equation}
where $R^* = \lambda \, R/a_V$ is a normalized radial coordinate in the intermediate configuration.
While this second-order approximation is not an exact solution for the finite-amplitude translated sphere problem, it is a standard continuation of the linear elastic solution to finite deformation.
As detailed in \ref{app:translate-sphere-second-order}, the approximate solution satisfies incompressibility to the order of $\eta^2$ and, for a homogeneous neo-Hookean solid, mechanical equilibrium is also satisfied to the order of $\eta^2$.

\subsection{Rayleigh--Ritz minimization of strain energy}
\label{subsec:ro-second}

For the unbounded body containing an inflated-and-translated spherical cavity, we assume that the deformation gradient can be approximated in the multiplicative decomposition form shown in \cref{eqn:F_coupled}, with $\defgrad^{\Lambda}$ and $\defgrad^{\eta}$ calculated according to \cref{eqn:F_infl} and \cref{eqn:F_trans}, respectively.
Following an approach similar to our first-order model, we seek $\eta$ that globally minimizes the strain energy for a graded material when $\Lambda_V$ is prescribed. 

For an incompressible neo-Hookean material with $J = 1$, the strain energy density \cref{eqn:nh-strain-energy} is simplified as
\begin{equation}
    \psi = \frac{G\(R,\Theta\)}{2} \tr{\defgrad \, \defgrad^{\transp} - 3}
    \, .
\end{equation}
Integrating $\partial{\psi}/\partial{\eta}$ over the unbounded body, similar to the step leading to \cref{eqn:energy-min-condition}, we obtain
\begin{equation}
    \eta^{\rm RR2} = \alpha \, \Htwo \( L^*, \Lambda_V \) \(1 - \Lambda_V\)    
\, ,
\label{eqn:ro-2_eta}
\end{equation}
where 
\begin{equation}
\begin{aligned}
    \Htwo \( L^*, \Lambda_V \) &= \(\Lambda_V + 1 + \Lambda_V^{-1}\) \, {\driveQ}/{\stiffK}
    \, , \\
    \driveQ &= \int_{-1}^{1} \int_{-1}^{1}  \frac{ c \, \(\Lambda_V^3 - 1 - 16/\(\xi - 1\)^3\) }{\(\xi - 1\)^2 \(\Lambda_V^3 - 1 - 8/\(\xi - 1\)^3\) ^2} \tanh\( \frac{2 \, c}{L^*\(1-\xi\)} \) \dd{c} \, \dd{\xi}
    \, , \\
    \stiffK &= \int_{-1}^{1}  \frac{ 128 }{\(\xi - 1\)^8 \(\Lambda_V^3 - 1 - 8/\(\xi - 1\)^3\)^{8/3}}  \dd{\xi}
    \, .
\label{eqn:ro2_H}
\end{aligned}
\end{equation}
Whereas $\stiffK$ depends only on $\Lambda_V$, $\driveQ$ varies with both $\Lambda_V$ and $L^*$.
The resulting dependence of $\eta^{\rm RR2}$ on $L^*$ is examined through the steep- and gradual-gradient limits below.

% ===============

\subsection{Asymptotic behaviors related to $L^*$}
\label{subsec:discuss_eta-asympt}

\begin{figure}[tb]
    \centering
    \includegraphics[width=0.75\linewidth]{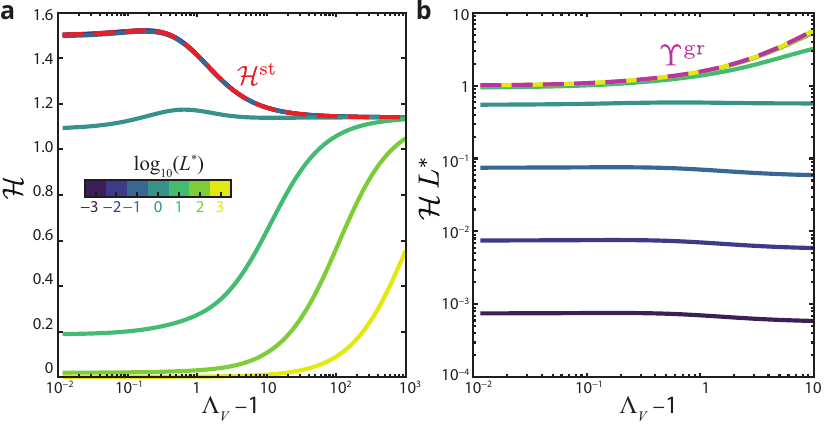}
    \caption{
    Asymptotic behaviors predicted by the second-order Rayleigh--Ritz model.
    (a) As the stiffness gradient becomes steeper with decreasing $L^*$, $\Htwo$ approaches an asymptotic limit value, $\Htwo^{\rm st}$.
    (b) As the stiffness gradient becomes more gradual with increasing $L^*$, $\Htwo \, L^*$ approaches an asymptotic limit value, $\HL^{\rm gr}$. 
    }
    \label{fig:ro-2_asymptote}
\end{figure}

As the stiffness gradient approaches the steep limit, with $L^* \to 0$, the stiffness gradation profile resembles a step discontinuity and the hyperbolic tangent function appearing in $\driveQ$ converges to 
\begin{equation}
    \lim_{L^* \to 0} \( \tanh\( \frac{2 \, c}{L^* \(1 - \xi\)} \) \) = \sign\( c \) 
    \, ,
\label{eqn:ro2_steep-limits}
\end{equation}
independent from $L^*$.
Conversely, as the stiffness gradient approaches the gradual limit, with $L^* \to \infty$, 
we have
\begin{equation}
    \lim_{L^* \to \infty} \(L^* \, \tanh\( \frac{2 \, c}{L^* \(1 - \xi\)} \) \) = \frac{2 \, c}{1-\xi} 
    \, .
\label{eqn:ro2_gradual-limits}
\end{equation}
Accordingly, $\eta^{\rm RR2} \propto 1/L^*$ when $L^* \to \infty$. 
Therefore, the second-order Rayleigh--Ritz model recovers both asymptotic trends observed in \Cref{subsec:fe_eta-length} from finite-element simulations.

Bringing the results of \cref{eqn:ro2_steep-limits} and \cref{eqn:ro2_gradual-limits} into \cref{eqn:ro2_H}, we can numerically evaluate the asymptotic values related to the steep- and gradual-gradient limits,
\begin{equation}
    \Htwo^{\rm st}\(\Lambda_V\) = \lim_{L^* \to 0} \Htwo
    \, , \quad
    \HL^{\rm gr}\( \Lambda_V \) = \lim_{L^* \to \infty}\(\Htwo \, L^*\)
    \, .
\end{equation}

In \Cref{fig:ro-2_asymptote}~(a), we show $\Htwo$ for various normalized length scales $L^*$. 
As $L^*$ is decreased, we observe a convergence of $\Htwo$ to $\Htwo^{\rm st}$.
When $L^* \leq 10^{-1}$, $\Htwo$ becomes visually indistinguishable from $\Htwo^{\rm st}$ throughout the range of $\Lambda_V$ examined.
Moreover, at sufficiently large $\Lambda_V$, $\Htwo$ for all $L^*$ eventually converges to a plateau value of approximately 1.14.
This rationalizes the $L^*$-independent, large-inflation plateau of $\eta$ noted earlier in \Cref{subsec:fe_eta-length}.

In \Cref{fig:ro-2_asymptote}~(b), we show that $\Htwo \, L^*$ approaches $\HL^{\rm gr}$ as $L^*$ increases. 
For small inflation with $\Lambda_V \leq 2$, $\HL^{\rm gr}$ matches closely to $\Htwo \, L^*$ for $L^* \geq 10$.
At larger inflation levels, the $L^* = 10$ case exhibits a pronounced divergence of $\Htwo \, L^*$ from $\HL^{\rm gr}$, while $\Htwo \, L^*$ for more gradual stiffness gradients with $L^* \geq 100$ remains in good agreement with $\HL^{\rm gr}$ up to $\Lambda_V \leq 10$.

% ===============
\subsection{Improved finite-inflation prediction}
\label{subsec:ro2-validation}

\begin{figure}[tb]
    \centering
    \includegraphics[width=0.75\linewidth]{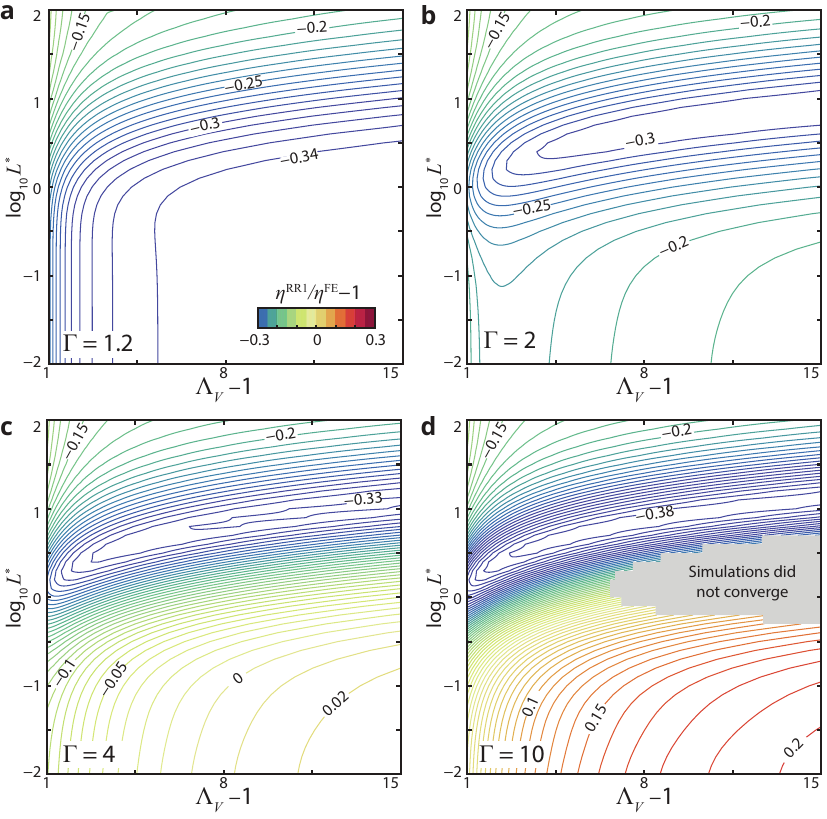}
    \caption{
    Signed relative error of the second-order Rayleigh--Ritz solution $\eta^{\rm RR2}$ compared to finite-element solution $\eta^{\rm FE}$, for representative cases with $\Zbar^* = 0$ and (a) $\Gamma = 1.2$, (b) $\Gamma = 2$, (c) $\Gamma = 4$, (d) $\Gamma = 10$.
    Consistent across various levels of stiffness contrast, the second-order model underpredicts the migration ratio for a gradual stiffness gradient with $L^* > 1$.
    For steep stiffness gradient with $L^* < 1$, the second-order model generally underpredicts cavity migration for small $\Gamma$ and overpredicts cavity migration for large $\Gamma$.
    For large stiffness contrast (e.g., $\Gamma = 10$) and $L^*$ on the order of unity, finite-element simulations did not robustly converge beyond a case-dependent limit of $\Lambda_V$.
    }
    \label{fig:ro-2_eta-evol}
\end{figure}

We now assess the capability of the second-order Rayleigh--Ritz model to quantify $\eta$ when $\Zbar^* = 0$.
In \Cref{fig:ro-2_eta-evol}, we present the signed relative error $\eta^{\rm RR2}/\eta^{\rm FE} - 1$ for the representative cases of $\Gamma = 1.2$, $2$, $4$, and $10$.
Compared to the first-order model presented earlier, the second-order model exhibits much improved accuracy for finite inflation.
For a very gradual stiffness gradient with $L^* = 100$, the second-order model underpredicts the migration ratio, with a consistent relative error magnitude of approximately $11\%$ at $\Lambda_V = 2$ across different levels of stiffness contrast.
The accuracy does not deteriorate substantially as $\Lambda_V$ further increases, with the relative error magnitude staying within $21\%$ at $\Lambda_V = 16$.

In contrast, the relative error for $L^* \ll 1$ is more sensitive to $\Gamma$, especially when $\Lambda_V$ is large and $\eta$ plateaus. 
The second-order model generally underpredicts the plateau value for $\Gamma \lesssim 4$ and overpredicts the plateau value for larger $\Gamma$.
Nonetheless, for $\Gamma \in \[2, 10\]$, the relative error magnitude of $\eta$ generally stays within $22 \%$ up to $\Lambda_V = 16$, when $\eta$ approaches the large-inflation plateau.
We have also confirmed that the magnitude of relative error between $\eta^{\rm RR2}$ and $\eta^{\rm FE}$ for $L^* = 1/100$ remained within $28 \%$ when $\Gamma$ is increased up to $20$ and $\alpha$ gradually saturates.

For $\Gamma \geq 2$, the disagreement between $\eta^{\rm RR2}$ and $\eta^{\rm FE}$ is generally largest for $1 \leq L^* \leq 10$, with relative error magnitude already exceeding $30 \%$ at $\Lambda_V = 5$.
In this range of $L^*$, the gradation length scale is comparable to the undeformed cavity radius.
Thus, the shear modulus varies appreciably within the near-cavity-wall region that undergoes the largest deformation during inflation.
As mentioned in \Cref{subsec:fe-intro} and \ref{app:fem}, it is also near this range of moderate $L^*$ that our finite-element simulations struggled to proceed for large inflation levels when the stiffness contrast $\Gamma$ was high---for example, as shown in \Cref{fig:ro-2_eta-evol}~(d), simulations for $\Gamma = 10$ and $L^* = 1$ did not robustly converge beyond $\Lambda_V = 7.7$.
Perhaps, the deformation state in these cases may be more complex than the simplified form assumed in our reduced-order setting.

% =====================
\section{Discussion}
\label{sec:discussion}

% ====================

\subsection{Limitations of the trial kinematics}
\label{subsec:discuss_limitation}

\begin{figure}[tb]
    \centering
    \includegraphics[width=0.75\linewidth]{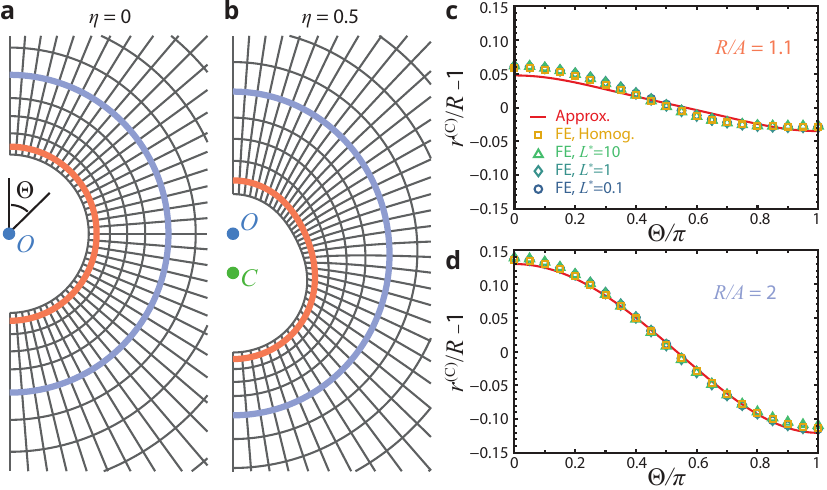}
    \caption{
    For second-order-accurate pure migration of cavity, displacement fields agree well between finite-element simulations and the analytical approximation used in the second-order Rayleigh--Ritz model.
    The accuracy is minimally sensitive to stiffness gradation along the direction of cavity migration (i.e., $-\basis_3$).
    Mesh of near-cavity domain for (a) $\eta = 0$ (i.e., undeformed) and (b) $\eta = 0.5$, according to the second-order analytical approximation. 
    The coarse mesh shown here is used only for visualization, while all finite-element calculations employed the refined discretization described in \ref{app:fem} and shown in \Cref{fig:mesh}.
    For the representative case of $\eta = 0.5$, the predicted $C$-centered radial coordinate change at constant-$R$ surfaces with (c) $R/A = 1.1$ and (d) $R/A = 2$ do not vary substantially between the analytical approximation, finite-element simulation in a homogeneous material, and finite-element simulations in graded materials with $\Gamma = 4$, $\Zbar^* = 0$, and $L^*$ ranging from $0.1$ to $10$.
    }
    \label{fig:fe-vs-ro_translate}
\end{figure}

\begin{figure}[tb]
    \centering
    \includegraphics[width=0.75\linewidth]{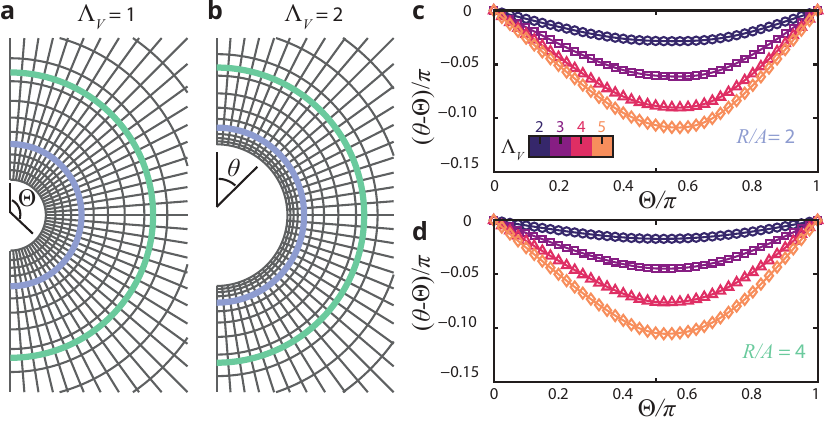}
    \caption{
    Finite-element simulations indicate that, when pure radial motion is imposed on cavity surface, stiffness gradation leads to polar displacement.
    Mesh of near-cavity domain for (a) $\Lambda_V = 1$ (i.e., undeformed) and (b) $\Lambda_V = 2$, according to displacement field assumed in the second-order Rayleigh--Ritz model. 
    Difference of current and reference polar coordinates are shown for constant-$R$ surfaces with (c) $R/A = 2$ and (d) $R/A = 4$, for the representative case of $\Gamma = 4, L^* = 1, \Zbar^* = 0$.
    As the volume-equivalent stretch $\Lambda_V$ increases, the polar angle change increases in magnitude. 
    }
    \label{fig:fe-vs-ro_inflate}
\end{figure}

Without involving empirical fitting parameters, the second-order Rayleigh--Ritz model predicts the primary trends in cavity migration throughout much of the finite-inflation regime examined.
Agreement with the finite-element simulations is particularly strong for sufficiently gradual stiffness gradients, with $L^* \gg 1$, and for sufficiently steep gradients, with $L^* \ll 1$, provided that the stiffness contrast is not too small, with $\Gamma \geq 2$.
In contrast, appreciable discrepancies remain when $L^*$ is on the order of unity or when $L^* \ll 1$ for $\Gamma < 2$.
The relatively poor accuracy in these parts of the parameter space likely arise from limitations of the trial deformation mappings assumed.

A crucial assumption in our Rayleigh--Ritz framework is that the deformation mappings for the radial inflation and the translation of a spherical cavity in a homogeneous material can be composed to approximate the inflation--migration response of a graded material.
Here, we present auxiliary finite-element simulations to assess this kinematic assumption.

\subsubsection{Validity of the translated sphere solution}

First, we test the second-order analytical approximation for shear-free translation of a spherical cavity, derived in \ref{app:translate-sphere-second-order}.
We reused the two-dimensional axisymmetric finite-element domain and discretization employed for the earlier cavity-inflation simulations, as described in \ref{app:fem}, but modified the boundary conditions.
In place of the fluid-cavity loading, we directly applied displacement boundary conditions to the inner surface of the hyperelastic body (i.e., nodes with reference radial coordinate $R = A$).
Specifically, the normal displacement prescribed by the analytical approximation was imposed at the boundary nodes, while the tangential displacement remained unconstrained, thereby approximating a sphere of radius $A$ migrating by a magnitude of $\eta \, A$ in a shear-free manner. 
To illustrate the assumed kinematics, we show in \Cref{fig:fe-vs-ro_translate}~(a) and (b), respectively, an undeformed mesh of the near-cavity domain and its image under the second-order analytical mapping at $\eta = 0.5$.

We first examined this displacement-controlled auxiliary problem in a homogeneous neo-Hookean material.
In \Cref{fig:fe-vs-ro_translate}~(c), we compare the second-order approximate and finite-element results for the current radial coordinate along the surface with $R/A = 1.1$ when $\eta = 0.5$.
The approximate solution agreed well with the homogeneous finite-element results in predicting a decrease of $r^{(C)}/R$ as the reference polar coordinate $\Theta$ increased. 
The relative error of the quantity $r^{(C)}/R - 1$ fell within $12\%$.
In \Cref{fig:fe-vs-ro_translate}~(d), we present a similar comparison between the second-order approximate and finite-element results along the surface with $R/A = 2$ when $\eta = 0.5$.
Although the angular variation in $r^{(C)}/R$ increased along this surface located farther away from the cavity, the relative error of the quantity $r^{(C)}/R - 1$ stayed within $2\%$ between the approximate solution and the homogeneous finite-element results.
We have also confirmed that the current polar coordinates calculated by the approximate solution agreed well with the homogeneous finite-element results along the constant-$R$ surfaces examined.

We repeated the same displacement-controlled auxiliary problem in graded neo-Hookean materials, considering $\Gamma = 4$, $\Zbar^* = 0$, and progressively steeper stiffness gradients with $L^* = 10$, $1$ and $1/10$.
Remarkably, the resulting displacement fields differed only minimally from the homogeneous-material solution.
This is illustrated in \Cref{fig:fe-vs-ro_translate}~(c) and (d) for the case of $\eta = 0.5$ and the representative constant-$R$ surfaces considered earlier.
The graded finite-element simulation results are indiscernible from those of the homogeneous finite-element simulation.
Introducing the stiffness gradation did not substantially increase the discrepancy between the finite-element solution and the second-order analytical approximation for the isolated migration mapping.
This is likely due to the alignment of the migration mode with the direction of stiffness gradation. 
Our results here suggest that the second-order approximate solution captures the migration-induced deformation sufficiently well, even in graded materials.

We reiterate that the second-order approximate solution ensures incompressibility to the order of $\eta^2$.
Throughout the unbounded body, the deviation of the Jacobian determinant of the deformation gradient, $J$, from unity is within a magnitude of $0.01$ for $\eta < 0.36$ and within $0.1$ for $\eta < 0.78$.
As $\eta$ increases, we naturally expect increased discrepancies between the displacement fields in the approximate solution and the finite-element simulation, even when material homogeneity is assumed. 
In principle, the remaining discrepancies can be reduced systematically by retaining higher-order terms in the migration expansion. 
However, such an extension would require solving the corresponding higher-order equilibrium and incompressibility equations and is not pursued here.

\subsubsection{Validity of the inflated sphere solution}

Next, we consider another set of auxiliary finite-element simulations in which a spherically symmetric radial displacement was prescribed at the inner boundary according to \cref{eqn:u_finite-inflation}, producing a cavity with volume-equivalent stretch of $\Lambda_V$.
We illustrate the deformation mapping in \Cref{fig:fe-vs-ro_inflate}~(a) and (b), showing an undeformed mesh and its deformed configuration at $\Lambda_V = 2$, respectively.
Since the assumed deformation mapping is an exact solution for cavity inflation in a homogeneous material, finite-element simulations in a homogeneous neo-Hookean body resulted in negligible differences between the assumed displacement field and the simulation results.

When this displacement-controlled auxiliary problem was repeated in a graded material, however, the finite-element results revealed gradation-induced adjustments that are absent from the homogeneous inflation mapping.
It appears that, due to the misalignment between the spherically-symmetric deformation assumed and the shear modulus gradation prescribed, an additional redistribution of displacement is needed to achieve mechanical equilibrium in the body.
With increasing inflation level, we observed a larger deviation of the current polar coordinate $\theta$ from the reference polar coordinate $\Theta$.%
\footnote{Since $O$ and $C$ coincide in this case, we denote the current polar coordinate directly as $\theta$.}
This is shown in \Cref{fig:fe-vs-ro_inflate}~(c) and (d) along surfaces with $R/A = 2$ and $4$, respectively, for the representative case of $\Gamma = 4$, $L^* = 1$, and $\Zbar^* = 0$.
Under the prescribed uniform radial displacement of the cavity surface, we expect the stiffer side of the material to develop larger stresses and strain-energy density near the cavity.
Consistent with this response, the imposed boundary condition produced larger radial displacements at smaller $\Theta$ on constant-$R$ surfaces.
The accompanying polar motion is directed toward the stiffer side, such that $\theta < \Theta$ for non-polar points, as shown in \Cref{fig:fe-vs-ro_inflate}~(c) and (d).

The gradation-induced angular variation of the radial displacement and the accompanying polar motion are absent from the homogeneous inflation mapping.
The omission of this displacement redistribution from the assumed kinematics may contribute to the error in the second-order Rayleigh--Ritz model.
The influence of this omission is expected to be particularly large when $L^*$ is on the order of unity and the shear modulus varies appreciably throughout the vicinity of the cavity.
This interpretation is consistent with our earlier observation in \Cref{subsec:ro2-validation} that, compared to $L^* \ll 1$ and $L^* \gg 1$ cases, a case with $L^*$ on the order of unity was captured less accurately by the second-order Rayleigh--Ritz model.
The contribution of the omitted redistribution is also expected to be more pronounced as the stiffness contrast $\Gamma$ increases.
A contrast-dependent redistribution of the inflation-related displacement field may help explain why the dependence of $\eta$ on $\alpha$ is not fully captured by the present Rayleigh--Ritz framework, as inferred previously in \Cref{subsec:ro1_eta-symm} from the $|\Zbar^*|$-dependent asymmetry at finite inflation.
The contrast-dependent redistribution may also explain the increased sensitivity of the Rayleigh--Ritz model's accuracy to $\Gamma$ when $L^*$ decreases, as seen from \Cref{fig:ro-2_eta-evol}.

\subsubsection{Implications for Rayleigh--Ritz predictions}

Despite the noted limitation of the trial kinematics assumed in our reduced-order framework, the second-order Rayleigh--Ritz model remains quantitatively effective at predicting the migration of an inflated cavity in a graded hyperelastic solid.
It appears that the success of the Rayleigh--Ritz framework does not strictly rely on an accurate, local reconstruction of the deformation mapping.
This distinction is philosophically reminiscent of many classical reduced-order models, such as the Euler--Bernoulli beam theory and the Hertz contact model, which can predict global responses effectively despite simplification of local fields.
By seeking a global strain-energy-minimizing equilibrium solution, the present Rayleigh--Ritz framework captures the energy balance in the graded material without necessarily resolving pointwise mechanical equilibrium.

The identification of the omitted, inflation-related redistribution of the displacement field provides a clear future direction for refining the present reduced-order framework.
The redistribution may be introduced as an additional intermediate mapping between the inflation and migration modes, resulting in a three-part decomposition of the deformation gradient.
We could then proceed with the Rayleigh--Ritz energy minimization procedure to arrive at a more accurate approximation of the migration ratio.

However, incorporating this inflation-related redistribution is not a straightforward extension.
The required correction to the displacement field does not appear to match an obvious universal mapping function that also preserves incompressibility. 
We also anticipate the redistribution field to depend jointly on $\Lambda_V$, $L^*$, and $\Gamma$.
The construction and validation of the corresponding displacement field would require additional theoretical development beyond the present work.

% ===================
\subsection{Model assumptions and possible extensions}

We intentionally restricted our reduced-order models to only consider the inflation and migration modes strongly influencing the cavity evolution.
The framework can be systematically enriched by constructing admissible bulk displacement fields associated with second- and higher-degree Legendre modes and enforcing stationarity of the approximated total strain energy with respect to the additional mode amplitudes.
Whereas the two-mode framework presented herein offers a clear interpretation of the comparatively pronounced migration response of a pre-existing cavity in a graded material, a multi-mode extension could improve agreement between the reduced-order predictions and the finite-element solutions.
Specifically, such an extension would explicitly account for coupling between centroid migration and nonspherical distortion modes and may thereby improve predictions of $\eta$.
Another natural extension is to relax the axisymmetry assumption made in this work and represent cavity shape distortion with the full spherical-harmonic basis.
Related spherical-harmonic expansions have been utilized to study nonspherical perturbations and surface instabilities of microbubbles during inertial cavitation in viscoelastic soft materials~\cite{gaudron-etal_2020_shape-stability,remillard-etal_2026_surface-instabilities}.
The Legendre modes used here constitute the axisymmetric subset of the spherical-harmonic basis.

To clarify the contribution of the graded elastic shear modulus, we intentionally neglected surface tension in our analysis.
We anticipate that surface tension would strengthen the relative prominence of cavity migration over nonspherical distortion, because capillarity penalizes nonspherical shape modes but does not penalize a rigid translation of a spherical surface. 
Still, it would be valuable to carefully quantify this effect in a future study, if primarily for micrometer-scale cavities resulting in capillary pressure comparable to the material's shear modulus~\cite{henann-bertoldi_2014_elasto-capillary,wang-henann_2016_liquid-inclusions}.

Although we consider a specific form of stiffness gradation profile in our investigation, our findings are relevant to a broader class of functionally graded materials.
In particular, the asymptotic conclusions for the steep- and gradual-gradient limits extend naturally to other gradation profiles described by $G\(Z\) = \Gbar\(1 + \alpha \, f\(\(Z - \Zbar\)/L\) \)$, where $f(x)$ is smooth and monotone, satisfies $f(0) = 0$ and $f'(0) \neq 0$, and converges to $\pm 1$ as $x$ approaches $\pm \infty$.
When the stiffness gradient is sufficiently steep relative to a pre-exiting cavity, we may approximate the gradation profile as a bi-material interface, consistent with \cref{eqn:ro2_steep-limits}.
When the stiffness gradient is sufficiently gradual, the gradation profile near the cavity may be approximated according to the leading-order term in the Taylor expansion of $G\(Z^*\)$.
As shown in \cref{eqn:ro2_gradual-limits}, this leads to $\eta \propto 1/L^*$ in the case of $f\(x\) = \tanh\(x\)$ and other common sigmoidal profiles such as $f\(x\) = {\rm erf}\(x\)$ and $f\(x\) = 2 \, {\arctan}\(x\)/\pi$.
However, when the gradation length scale is comparable to the undeformed cavity radius, the response may depend more strongly on the particular choice of $f(x)$, warranting a dedicated, profile-dependent analysis.

We also observed that a migration response, albeit a weak one, is present even when the cavity samples a nearly homogeneous neighborhood, either with $L^* \gg 1$ or $|\Zbar^*| \gg L^*$.
Intuitively, this seems to contradict the assumption of local homogeneity serving as the cornerstone of cavitation rheometry techniques to probe local mechanical properties.
However, we believe that our analysis serves to clarify the length-scale requirement underlying the locally homogeneous assumption and further strengthens the use of small cavities to probe the neighborhood of heterogeneous materials.
Measurable centroid migration does not, by itself, invalidate the assumption of local homogeneity and the corresponding inverse characterization results based on the dominant spherical inflation response.
Rather, a weak migration serves as a sensitive indicator of global stiffness gradation and further justifies the process of probing heterogeneous materials in a neighborhood-by-neighborhood manner.

% ===============

\subsection{A roadmap for parameter recovery}
\label{subsec:param-recovery}

We have established that the Rayleigh--Ritz framework reproduces the following trends distinctively relating the migration ratio $\eta$ to the gradation parameters:
\begin{enumerate}[label=(\Roman*)]
    \item the gradation midplane $Z^* = \Zbar^*$ serves as a plane of symmetry for the initial small-inflation growth rate of $\eta$;\label{item:eta_symm}
    \item for centered cavity with $\Zbar^* = 0$, the evolution of $\eta$ with $\Lambda_V$ exhibits an asymptotic behavior when $L^* \ll 1$; \label{item:eta_steep}
    \item for centered cavity with $\Zbar^* = 0$, the evolution of $\eta \, L^*$ with $\Lambda_V$ exhibits an asymptotic behavior when $L^* \gg 1$.   \label{item:eta_gradual}
\end{enumerate}

These parameter dependencies provide a potential basis for the inverse characterization of the graded hyperelastic material via cavity inflation experiments.
In principle, measurements of $\eta\(\Lambda_V\)$ for cavities with different initial positions and radii could be combined to estimate the parameters describing the graded material in the specimen-fixed frame: $\Gamma$, $L$, and $\refZbar$.
We therefore propose the following roadmap for sequential recovery of these material parameters:

\begin{tcolorbox}[
    colback=gray!5,
    colframe=gray!90,
    title={Parameter recovery from cavity inflation},
    fonttitle=\bfseries,
    boxrule=0.5pt,
    arc=2pt
]

\begin{enumerate}
    \item 
    \emph{Location of gradation midplane, $\refZbar$:} 
    scan along $\basis_3$, nucleate and inflate cavity of similar undeformed radius $A$ at different planes of $\refZ$.
    The gradation midplane with $\refZ = \refZbar$ is qualitatively identified from the symmetry of $\eta$ with respect to $\Zbar^*$ during the initial, small-inflation stage. 
    The direction of stiffening (i.e., $\basis_3$ for $\Gamma > 1$) can be confirmed according to the direction of centroid migration.

    \item 
    \emph{Stiffness contrast, $\Gamma$:} 
    At $\Zbar^* = 0$, inflate cavity with varying initial radius $A$. 
    For larger $A$, corresponding to smaller $L^*$, the quantity $\eta/\Htwo^{\rm st}$ collapses to a linear relation with $1 - 1/\Lambda_V$, with a slope of $\alpha = \(\Gamma - 1\)/\(\Gamma + 1\)$.
    
    \item 
    \emph{Gradation length scale, $L$:} 
    For smaller $A$ experiments at $\Zbar^* = 0$, corresponding to larger $L^*$, the quantity $\eta/\(A \, \HL^{\rm gr}\)$ collapses to a linear relation with $1 - 1/\Lambda_V$, with a slope of $\alpha/L$. 
    With $\alpha$ determined in the previous step, we find $L$.
\end{enumerate}

\end{tcolorbox}

To assess the feasibility of this parameter recovery procedure, we now use the finite-element results from \Cref{sec:fe-results} as noise-free synthetic measurements of $\eta\(\Lambda_V\)$.
Because the approximate symmetry of $\eta$ about $\Zbar = 0$ is already demonstrated in the simulations, the recovery of gradation midplane cannot be meaningfully tested with synthetic data.
Therefore, we directly assume that the gradation midplane is accurately identified and proceed to assess the recovery of $L$ and $\Gamma$, which relies on the quantitative predictability of the second-order Rayleigh--Ritz model. 

\begin{figure}[tb]
    \centering
    \includegraphics[width=0.5\linewidth]{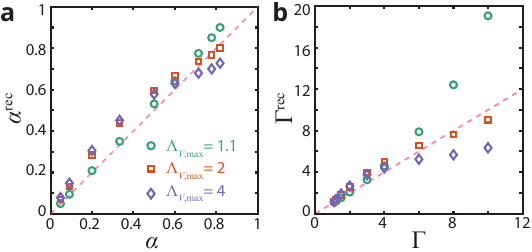}
    \caption{
    Target vs. recovered values of (a) gradation amplitude $\alpha$ and (b) stiffness contrast $\Gamma$, using the proposed parameter recovery procedure.
    The pink dashed line corresponds to a perfect agreement between the target and recovered values. 
    Using the second-order Rayleigh--Ritz model and synthetic measurements of $\eta\(\Lambda_V\)$ for a cavity with $A = 100 \, L$, a stiffness contrast of $\Gamma \leq 4$ can be recovered well.
    However, a larger stiffness contrast corresponds to a saturation of $\alpha$ and a poor recovery of $\Gamma$.
    }
    \label{fig:param-recovery}
\end{figure}

We recover $\alpha$ by considering the synthetic measurement corresponding to a large undeformed radius $A$ and then performing linear regression of $\(1 - 1/\Lambda_{V}\)$ vs. $\eta/\Htwo^{\rm st}$. 
\Cref{tab:inverse-sol} shows the results when $A = 100 \, L$ (i.e., $L^* = 1/100$) is used.
When only small-inflation data with $\Lambda_V \in \[1, 1.1\]$ is considered, $\alpha$ is generally recovered with relative error within $10 \%$.
The accuracy arises from the excellent agreement between the finite-element and Rayleigh--Ritz predictions of $\eta$ near $\Lambda_V = 1$.
The recovered $\alpha$ translates to a relatively accurate recovery of moderate stiffness contrast with $\Gamma \leq 4$, as illustrated in \Cref{fig:param-recovery}.
Due to the saturation of $\alpha$ for large $\Gamma$, the recovery of $\Gamma$ becomes ill-conditioned for large stiffness contrast.
If we were to use $A = 10 \, L$ (i.e., $L^* = 1/10$) for the large-cavity measurement, the accuracy of recovering $\Gamma \leq 4$ is minimally affected.

However, changes in cavity geometry are challenging to quantify experimentally when changes in $\Lambda_V$ are small.
A larger range of inflation level should be considered for a more practical inverse characterization procedure.
For the cases of $\Lambda_V \in \[1, 2\]$ and $\Lambda_V \in \[1, 4\]$ shown in \Cref{tab:inverse-sol} and \Cref{fig:param-recovery}, we find that the expanded inflation regime generally leads to a decline in the accuracy of the recovered stiffness contrast when $\Gamma \leq 4$, with a relative error up to $33\%$ from the target values. 
For larger $\Gamma$, expanding the range of $\Lambda_V$ may lead to a small improvement in the recovery of stiffness contrast, because the Rayleigh--Ritz model transitions from underestimating to overestimating $\eta$ in the moderate-$\Lambda_V$ inflation regime.

\begin{table}[tb]
\caption{Recovered values of $\Gamma$ and $\alpha$ using $\eta$ from finite-element simulations.
    }
\centering

\begin{tabular}{ c c | c c c c c c c c c}
   \multirow{2}{*}{Target}   & $\Gamma$ &  $1.1$ & $1.2$ & $1.5$ & $2$ & $3$ & $4$ & $6$ & $8$ & $10$  \\
                             & $\alpha$ &  $0.0476$ & $0.0909$ & $0.2$ & $0.333$ & $0.5$ & $0.6$ & $0.714$ & $0.778$ & $0.818$  \\ 
\hline
    \multirow{2}{*}{$\Lambda_{V,\max} = 1.1$} & $\Gamma$ & $1.10$ & $1.21$ & $1.53$ & $2.08$ & $3.27$ & $4.60$ & $7.89$ & $12.42$  & $19.07$  \\  
                            & $\alpha$ & $0.0496$ & $0.0948$ & $0.209$ & $0.350$ & $0.531$ & $0.643$ & $0.775$ & $0.851$ & $0.900$    \\
  \multirow{2}{*}{$\Lambda_{V,\max} = 2$} & $\Gamma$ & $1.15$ & $1.31$ & $1.80$ & $2.56$ & $3.95$ & $5.01$ & $6.58$ & $7.62$  & $9.05$  \\  
                            & $\alpha$ & $0.0701$ & $0.133$ & $0.284$ & $0.439$ & $0.596$ & $0.667$ & $0.736$ & $0.768$ & $0.801$   \\
  \multirow{2}{*}{$\Lambda_{V,\max} = 4$} & $\Gamma$ & $1.17$ & $1.35$ & $1.88$ & $2.65$ & $3.75$ & $4.41$ & $5.25$ & $5.67$  & $6.35$  \\  
                            & $\alpha$ & $0.0785$ & $0.148$ & $0.306$ & $0.452$ & $0.579$ & $0.630$ & $0.680$ & $0.700$ & $0.728$    
\end{tabular}

\label{tab:inverse-sol}
\end{table}

The recovery of $L$ involves another linear regression process, between $\(1 - 1/\Lambda_V\)$ and $\Htwo/\(A \, \HL^{\rm gr}\)$, for a synthetic measurement corresponding to a small undeformed radius $A$.
When $A = L/100$ (i.e., $L^* = 100$) is used, we generally underestimate $1/L$ with a relative error within $0.2\%$ when using $\Lambda_V \in \[1, 1.1\]$ and over-estimate by a relative error within $17\%$ when using $\Lambda_V \in \[1,2\]$ and $20\%$ when using $\Lambda_V \in \[1, 4\]$.
This level of error is consistent across $\Gamma \in \[1.1, 10\]$.
If we use $A = L/10$ (i.e., $L^* = 10$) as the small-cavity measurement, $1/L$ is underestimated by a relative error within $6\%$ for $\Lambda_V \in \[1, 1.1\]$ and overestimated by $10\%$ for $\Lambda_V \in \[1,2\]$ and $13\%$ for $\Lambda_V \in \[1,4\]$.
We note that these results assume a perfect recovery of $\alpha$ in the previous step.

These preliminary results establish the feasibility of identifying $L$ and moderate $\Gamma$ within an idealized, noise-free setting.
Varying the initial radii $A$ while maintaining $\Zbar = 0$ allows the same material profile to be sampled between the steep- and gradual-gradient regimes through $L^* = L/A$.
However, in order to extend this analytically simple roadmap to an applicable material characterization protocol, many practical issues must be addressed.
An inherent assumption in the illustrative examples above is that, cavities with radii spanning two to four orders of magnitude can be generated with the same experimental setup, thereby accessing both the steep-gradient limit, $\Htwo \to \Htwo^{\rm st}$ as $L^* \to 0$, and the gradual-gradient limit, $\Htwo \, L^* \to \HL^{\rm gr}$ as $L^* \to \infty$, of the second-order Rayleigh--Ritz model.
Spanning such a disparate range of cavity size may be challenging in practice.
The cavity migration also becomes difficult to measure in the gradual-gradient limit, where a small $L^* = L/A$ implies that the cavity radius is much smaller than the gradation length scale.
Since $\eta \propto 1/L^* = A/L$ in the gradual-gradient limit, the dimensional migration amplitude scales as $\delta \propto A^2/L$.
With decreasing $A$, the corresponding migration amplitude may become increasingly challenging to distinguish from measurement noise, which in turn limits the accuracy of the parameter recovery procedure.
More broadly, the development of an inverse characterization procedure requires an extensive set of tests to quantify the sensitivity of the procedure to measurement noise and parameter non-uniqueness~\cite{kaipio-somersal_2007_statistical-inverse-problems,jones-etal_2018_parameter-non-uniqueness,nikolov-etal_2026_sbvf}.
These tests are not trivial and, therefore, left for future work.

% ====================

\subsection{Connection to interfacial inertial cavitation}

The present work considers the quasi-static inflation of a pre-existing cavity in an incompressible hyperelastic solid whose shear modulus varies monotonically along a reference Cartesian direction.
While previous studies of this problem has been limited, a valuable point of comparison is the evolution of an inertial cavitation bubble in a fluid near a free surface or a solid boundary, a topic that has been more extensively investigated~\cite{supponen-etal_2016_scaling-single-bubble-jets, brujan-etal_2004_non-newtonian-rigid-boundary, sieber-etal_2023_cavitation-near-tissue-mimic}.
In such dynamic problems, the most prominent outcomes of the boundary-induced asymmetry are associated with the violent collapse stage of the bubble, during which rapid motion produces jetting and severe nonspherical distortion of the bubble.
However, experimental visualizations also suggest that the spherical symmetry is broken during the bubble growth stage: the bubble centroid can migrate toward the less confining side of the interface and the bubble geometry becomes mildly nonspherical~\cite{supponen-etal_2016_scaling-single-bubble-jets}.
A similar growth-stage trend was reported in a recent experimental study of cavitation bubbles nucleated on either side of a hydrogel--water interface~\cite{yang-etal_2026_interface-cavitation}.
These observations are qualitatively consistent with our findings: in the graded materials examined herein, quasi-statically inflated cavities migrate toward the more compliant side of the material while remaining nearly spherical.

We acknowledge that the analogy is qualitative, as the underlying mechanisms differ between the interfacial inertial cavitation problem and the present problem.
In the former, migration and distortion are governed by transient fluid motion and a sharp transition not only in shear modulus but also in density, compressibility, and acoustic impedance.
In our quasi-static analysis, inertia is absent and the asymmetry of cavity geometry is caused by a smooth gradation of elastic shear modulus.
Nonetheless, the relative prominence of centroid migration compared with nonspherical distortion in both settings suggests a useful common viewpoint: in a heterogeneous medium, the growth of a quasi-statically inflated cavity and an inertial cavitation bubble may both be fruitfully decomposed into centroid migration and recentered shape modes.
Although our Rayleigh--Ritz framework is based on elastic energy minimization in a quasi-static setting, the decomposition of cavity evolution into inflation, centroid migration, and nonspherical shape modes could guide future reduced-order analysis of inertial cavitation. 

Overall, this work highlights that, for a cavity inflated in a hyperelastic solid with shear modulus increasing along a reference Cartesian direction, centroid migration is the most prominent geometric signature of the stiffness gradation.
The proposed Rayleigh--Ritz framework analytically describes the effects of finite inflation and stiffness gradation on the migration ratio and agrees well with finite-element simulation results over a broad range of parameters examined. 
The reduced-order framework also reproduces the leading-order positional symmetry at small inflation and the asymptotic behaviors associated with extremely steep and gradual stiffness gradients, thereby clarifying the respective roles of the gradation midplane location, the stiffness contrast, and the gradation length scale.
These results establish a mechanics-based foundation for the inverse characterization of graded hyperelastic materials and reveal a broadly applicable strategy for the reduced-order analysis of cavity growth and bubble dynamics in heterogeneous media.

% =====================
%\section{Conclusions}
%\label{sec:conclusions}

% =====================
% Statements
% =====================
\section*{Conflicts of interest}

The authors have no known conflicts of interest associated with this publication, and there has been no significant financial support for this work that could have influenced its outcome.

\section*{Declaration of generative AI and AI-assisted technologies in the manuscript preparation process}

During the development of this work, the authors used U-M GPT, the University of Michigan’s generative-AI platform, to support brainstorming, preliminary verification of algebraic derivations, and manuscript editing. 
Generative AI was not used to generate or analyze the finite-element data reported in this study.
All mathematical results and written content assisted by the AI tool were independently reviewed and revised as needed by the authors, who take full responsibility for the manuscript.

%% ==================================
%% APPENDIX ========================
\appendix
% =====================

\section{Finite element model}
\label{app:fem}

\begin{figure}[tb]
    \centering
    \includegraphics[width=0.75\linewidth]{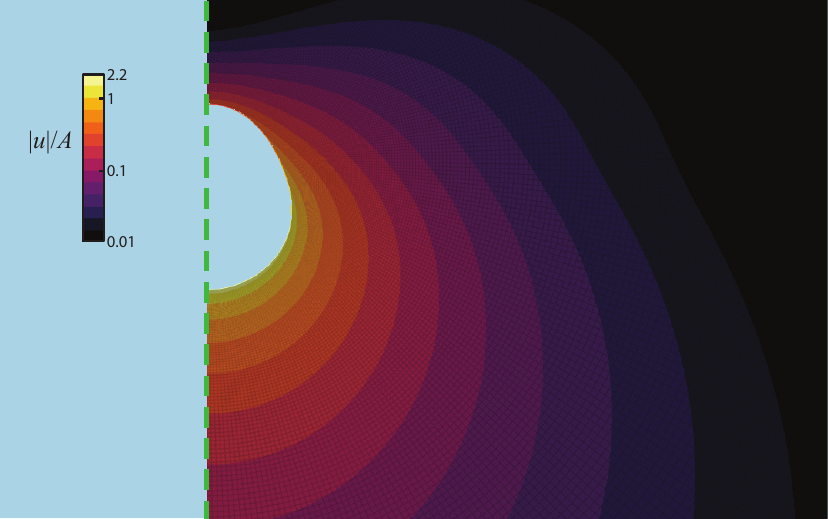}
    \caption{
    An illustration of the two-dimensional axisymmetric finite-element configuration.
    Deformed mesh overlaying normalized displacement magnitude, $|\bu|/A$, for $\Gamma = 4, L^* = 1, \Zbar^* = 0$.
    The more compliant, $\Theta = \pi$, end of the cavity is displaced more than the stiffer, $\Theta = 0$, end.
    The green dashed line corresponds to the axis of symmetry for the axisymmetric finite-element model.
    }
    \label{fig:mesh}
\end{figure}

Given the axisymmetric nature of the cavity inflation problem, our finite-element model used the {\tt{CAX4H}} (four-node, two-dimensional axisymmetric, constant-pressure) elements, as illustrated in \Cref{fig:mesh}.
In the stress-free reference configuration, the simulation domain was divided into 200 uniform increments along the polar direction $\Theta$ and the far-field radius was set to be $R_{\max} = 1000~A$.
We verified that further refinement of the mesh and increase of $R_{\max}$ have negligible effects on quantities of interest, such as pressure--volume response, migration amplitude, and recentered shape modes.
The far-field boundary nodes had negligible displacements as the cavity was inflated, thereby approximating an infinite domain.
To remove rigid-body motion of the domain, the node corresponding to $R = R_{\max}$ and $\Theta = 0$ was constrained along the $\basis_3$ direction (i.e., ${\tt{U2}} = 0$ in Abaqus syntax).

Because of our primary interest in the evolution of cavity geometry as a function of the volume-equivalent stretch $\Lambda_V$, we used a volume-controlled loading scheme, rather than prescribing pressure.
Specifically, we followed the procedure presented by \citet{kolluri_2025_thin-IMR} and prescribed mass addition into an isothermal fluid cavity.
This procedure avoids the poor conditioning of a pressure-control scheme near limiting or non-monotonic pressure--volume response, where small changes in pressure can produce large changes in cavity volume or make parts of the equilibrium path inaccessible.

To retain compatibility with the fluid-cavity loading framework, the surrounding material was modeled to be nearly incompressible, with bulk modulus $K = 10^5~\Gbar$.
We verified that further increase of $K$ has negligible effects on quantities of interest.
The spatial gradation of shear modulus was achieved with a {\tt{SDVINI}} subroutine that recorded the reference coordinate $Z$ of integration points as a state variable and a {\tt{UHYPER}} subroutine that calculates the strain energy density and its derivatives according to the local deformation state. 
As shown in \cref{eqn:nh-strain-energy}, the strain energy density is defined in a neo-Hookean form.

We note that, in cases with high stiffness contrast $\Gamma$ and a normalized gradation length $L^*$ near unity, the finite-element simulations sometimes became numerically challenging.
When the inflation level $\Lambda_V$ became sufficiently large in these case, the default convergence criteria in Abaqus/Standard were not met after repeated automatic cutbacks.
The difficulty appears to be associated with the delicate, coupled nature of the mass-addition loading framework: at each loading increment, the cavity mass is advanced and the solver must find an updated displacement field in the graded solid together with the corresponding cavity volume and cavity pressure.
For loading increments that failed to converge, solver logs indicated displacement or pressure corrections that remained large relative to the attempted increments.
The resolution of these numerical challenges may require dedicated strategies for continuation, stabilization, or remeshing, which were not pursued here.

% =====================
\section{Modal analysis of additional representative cases}
\label{app:more-fe-results}

\begin{figure}[tb]
    \centering
    \includegraphics[width=0.75\linewidth]{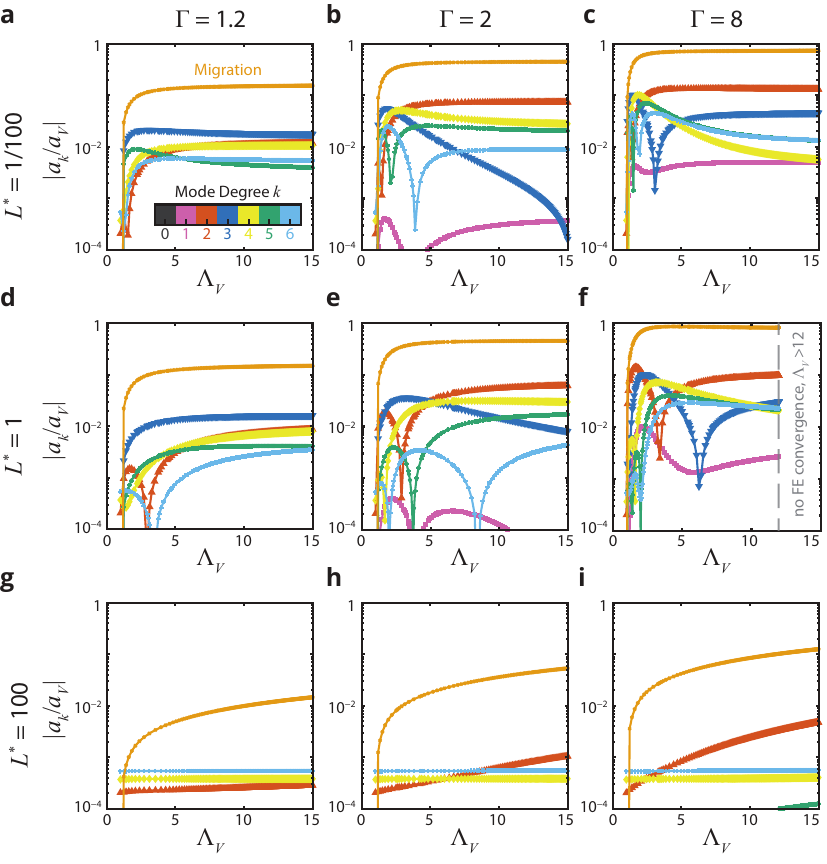}
    \caption{
    Finite-element simulation results for additional representative cases with $\Zbar^* = 0$, showing absolute value of mode amplitude normalized by current volume-equivalent radius $a_V$.
    For the purpose of comparison, we denote $a_k = \delta$ for the centroid migration.
    (a) $\Gamma = 1.2$, $L^* = 1/100$.
    (b) $\Gamma = 2$, $L^* = 1/100$.
    (c) $\Gamma = 8$, $L^* = 1/100$.
    (d) $\Gamma = 1.2$, $L^* = 1$.
    (e) $\Gamma = 2$, $L^* = 1$.
    (f) $\Gamma = 8$, $L^* = 1$; the finite-element simulation failed to robustly converge when $\Lambda_V > 12$ in this case.
    (g) $\Gamma = 1.2$, $L^* = 100$.
    (h) $\Gamma = 2$, $L^* = 100$.
    (i) $\Gamma = 8$, $L^* = 100$.
    }
    \label{fig:fe-extra}
\end{figure}

\begin{figure}[tb]
    \centering
    \includegraphics[width=0.5\linewidth]{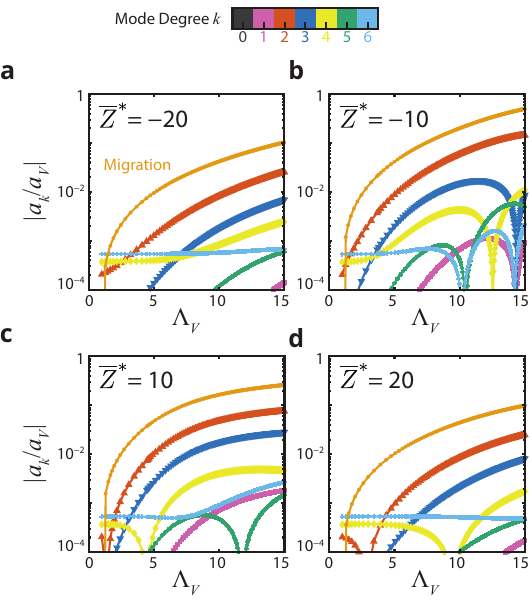}
    \caption{
    Finite-element simulation results for additional representative cases with $\Gamma = 4$ and $L^* = 1$, showing absolute value of mode amplitude normalized by current volume-equivalent radius $a_V$.
    For the purpose of comparison, we denote $a_k = \delta$ for the centroid migration.
    (a) $\Zbar^* = -20$.
    (b) $\Zbar^* = -10$.
    (c) $\Zbar^* = 10$.
    (d) $\Zbar^* = 20$..
    }
    \label{fig:fe-extra2}
\end{figure}

Here, we present additional finite-element simulation results to supplement the modal analysis presented in \Cref{subsec:fe_mode-evol}.
\Cref{fig:fe-extra} shows the evolution of centroid migration and nonspherical distortion mode amplitudes for gradation-centered cavities ($\Zbar^* = 0$) with stiffness contrast $\Gamma$ of $1.2$, $2$, and $8$ and normalized gradation length $L^*$ of $1/100$, $1$, and $100$.
In an initial range of small $\Lambda_V$, the modal amplitude stayed within $\pm a_V/100$, corresponding to negligible migration and nonspherical distortion. 
At larger inflation levels, the migration amplitude $\delta$ generally remained several times larger than that of the $k = 2$, the strongest nonspherical distortion mode.
In \Cref{fig:fe-extra2}, we present cases with $\Gamma = 4$ and $L^* = 1$, while $\Zbar^*$ varies from $-20$ to $20$.
As the cavity is offset from the gradation midplane, the onset of appreciable migration and nonspherical distortion are delayed, with amplitude remaining small for a larger initial range of $\Lambda_V$ than the $\Zbar^* = 0$ case shown in \Cref{fig:fe-raw}.
However, beyond this initial range, the migration amplitude is clearly larger than the amplitude of any nonspherical distortion mode.
These results are consistent with our findings from \Cref{fig:fe-raw}~(b) and confirm that migration is the most prominent form of symmetry-breaking response across the range of parameters examined.

% =====================

\section{Linear elastic solution of the translated sphere problem}
\label{app:translate-sphere-soln}

We consider a scalar potential $\chi$ and a vector potential $\bvphi$ that together allow us to express the displacement $\bu$ at material point $\bX$ in the Papkovich--Neuber form~\cite{sadd_2025_elasticity},
\begin{equation}
     \bu = 4 \(1 - \nu\) \bvphi - \Grad{ \chi + {\bvphi \cdot \bX} } 
\, ,
\label{eqn:u_potential}
\end{equation}
where $\Grad{\bullet}$ is the material gradient, taken with respect to the reference configuration.
Although the distinction between the reference and current configurations are commonly neglected in a small-deformation, linear-elasticity setting, we wish to maintain consistency with the finite-deformation problem considered in the main text and \ref{app:translate-sphere-second-order}.

For a homogeneous, isotropic material with shear modulus $G$ and Poisson's ratio $\nu$, the balance of linear momentum leads to
\begin{equation}
     \Lapl{\bu} + \frac{1}{1-2\nu} \Grad{\Div{\bu}} = -\frac{\bb}{G}
\, ,
\end{equation}
where $\Div{\bullet}$ is the material divergence.

Here, we are interested in an unbounded body containing a rigid sphere that is displaced by $\hat{\bu} = -\delta \, \basis_3$.
The corresponding potential functions are
\begin{equation}
     \bvphi = {\blob}_{\varphi} \frac{\hat{\bu}}{R}
\, , \quad
      \chi = {\blob}_{\chi} \, \frac{\hat{\bu} \cdot \bX}{R^3}
\, ,
\end{equation}
where $\blob_\varphi$ and $\blob_\chi$ are scalar constants to be determined.
% Self note: \chi fundamentally has the form (scalar)*(R^{-2})*P_1(\cos[\Theta]). 
% This matches the general form of solution mentioned in AWH Page 726.

Imposing incompressibility with $\nu = 1/2$, we can then write the components of the displacement field (in $O$-centered spherical coordinate frame) as
\begin{equation}
   u_R = \frac{2 \, \cos\Theta}{R^3}\(\blob_\chi + \blob_\varphi R^2\) 
   \, , \quad
    u_{\Theta} \frac{\sin\Theta}{R^3}\(\blob_\chi - \blob_\varphi R^2\)
    \, , \quad
    u_{\Phi} = 0
\, .
\end{equation}

In our cavity inflation problem, we expect traction to be purely normal at the cavity surface.
To match this condition, we require $\epsilon_{R\Theta} = \(R \, \partial{u_{\Theta}}/\partial{R} + \partial{u_{R}}/\partial{\Theta} - u_{\Theta}\)/\(2 R\) = 0$ for $R = A$.
Additionally, we require $u_R = -\delta \, \cos\( \Theta \)$ for $R = A$, conforming to the translated sphere. 
Incorporating these boundary conditions, we obtain
\begin{equation}
    {\blob}_{\varphi} = - \frac{\delta \, A}{2}
\, , \quad
    {\blob}_{\chi} = 0
\, ,
\end{equation}
and accordingly
 \begin{equation}
      u_R = - \frac{\delta \, A}{R} \cos\Theta
\, , \quad
      u_\Theta = \frac{\delta \, A}{2R} \sin\Theta
\, .
\label{eqn:translate-sphere-displ}
\end{equation}
This result is, perhaps, better known in the fluid mechanics community. 
It is a special case of the Hadamard–-Rybczynski solution, describing a creeping flow around a rigid sphere~\cite{leal_2007_transport}. 

To aid the second-order solution developed in \ref{app:translate-sphere-second-order}, we shall also solve for the strain and stress components in the unbounded body. 
The displacement field \cref{eqn:translate-sphere-displ} results in the infinitesimal strain $\beps$  with components
\begin{equation}
    \epsilon_{RR} = \frac{A \, \delta \, \cos\Theta}{R^2}
    \, , \quad
    \epsilon_{\Theta\Theta} = \epsilon_{\Phi\Phi} = -\frac{A \, \delta \, \cos\Theta}{2R^2}
    \, , \quad
    \epsilon_{R\Theta} = \epsilon_{R\Phi} = \epsilon_{\Theta\Phi} = 0
    \, .
\end{equation}
The total stress is the sum of a deviatoric part, equal to $2 \, G \, \beps$, and a hydrostatic part, the magnitude of which we denote as $\sigma_{\rm h}\(R,\Theta\)$.
The balance of linear momentum in the radial direction leads to
\begin{equation}
\begin{aligned}
    \fracp{}{R}\(2 \, G \, \epsilon_{RR} + \sigma_{\rm h}\) + 
    \frac{2 \, G}{R}\(2 \, \epsilon_{RR} - \epsilon_{\Theta\Theta} - \epsilon_{\Phi\Phi} 
    \) &= 0  
     \\
\quad \Rightarrow \quad
      \fracp{\sigma_{\rm h}}{R} + \frac{2\, G \, \delta \, A}{R^3} \cos\Theta  &= 0
      \\
\quad \Rightarrow \quad
      \sigma_{\rm h} = \frac{G \, \delta \, A}{R^2} \cos\Theta &+ \blob_{\rm h} 
\, .
\end{aligned}
\end{equation}
Because we expect $\sigma_{\rm h}$ to vanish as $R \to \infty$, we have $\blob_{\rm h} = 0$.
Therefore, $\sigma_{\rm h} = \(G \, \delta \, A/ R^2\) \cos\Theta$.

% =====================

\section{Second-order solution of the translated sphere problem}
\label{app:translate-sphere-second-order}

We consider an unbounded body of homogeneous, incompressible, neo-Hookean material, containing a sphere of radius $a_V$.
The centroid of the sphere is translated away from its origin $O$ by $\hat{\bu} = - \eta \, a_V \, \basis_3$ in a shear-free manner.
In \Cref{sec:second-order}, this corresponds to a deformation mapping from the intermediate configuration, containing an inflated cavity, to the current configuration, containing an inflated-and-translated cavity.
However, for the purpose of developing a second-order approximate solution here, we treat the intermediate configuration as the reference configuration and use $\Grad{\bullet}$ to denote gradient with respect to this configuration containing a sphere of radius $a_V$ centered at $O$.
The reference radial and polar coordinates $R$ and $\Theta$ used in this section are equivalent to the intermediate coordinates $\hat{R}$ and $\hat{\Theta}$ introduced in \Cref{sec:second-order}.
Furthermore, we denote $R^* = {R}/a_V$ as the relative radial coordinate of material points.

As shown in \cref{eqn:series_u}, we adopt an approximate, series-expansion form of displacement field, $\bu^{\eta} = \eta \, \bv + \eta^2 \, \bw$, where $\bv$ is a normalized first-order displacement from our linear-elastic solution in \ref{app:translate-sphere-soln} and $\bw$ is a normalized second-order displacement to be determined.  
Following standard notation for small-on-large deformation problems~\cite{baek-etal_2007_small-on-large}, we have the deformation gradient
\begin{equation}
    \defgrad = \iden + \eta \, \onegrad + \eta^2 \, \twograd
    \, , 
    \label{eqn:second_def-grad}
\end{equation}
where $\onegrad = \Grad{\bv}$, $\twograd = \Grad{\bw}$, and $\iden$ is identity.%
\footnote{
Strictly speaking, $\defgrad$ is a two-point tensor.
As shown by \citet{marsden-hughes_1994_math-foundations-elasticity}, the Euclidean space allows us to write $\defgrad = \shifter \, \tilde{\defgrad}$, where $\shifter$ is the Euclidean shifter and $\tilde{\defgrad} = \iden + \eta \, \onegrad + \eta^2 \, \twograd$ is a material tensor.
Given the fixed Euclidean basis in our problem, the shifter $\shifter$ is a two-point identity map that may be represented by the identity matrix. 
For brevity, we suppress the identity map associated with $\shifter$ and effectively write $\defgrad = \tilde{\defgrad}$ in our derivation.
}
As in \ref{app:translate-sphere-soln}, we consider a $O$-centered frame in both the reference and current configurations to describe the displacement field.

From \cref{eqn:trans_u}, the components of $\onegrad$ are
\begin{equation}
    \begin{bmatrix} \onegrad \end{bmatrix} = 
    \frac{1}{R^{*2}}
\begin{bmatrix}
       \cos\Theta & 
       \sin\Theta/2 & 
       0    
\\
       -\sin\Theta/2 &  
       -\cos\Theta/2 & 
        0   
\\
       0 & 
       0 & 
      -\cos\Theta/2    
\end{bmatrix} 
\, .
\end{equation}

The Jacobian determinant of the deformation gradient may be written as 
\begin{equation}
    J = \det{\defgrad} = \exp\( \tr{\ln\(\defgrad\) } \)
      = 1 + \eta \, \tr{\onegrad} + \eta^2\(\tr{\twograd} + \frac{1}{2}\(\tr{\onegrad}\)^2 - \frac{1}{2}\tr{\onegrad\onegrad}\) + \bigo\(\eta^3\)
    \, ,
\end{equation}
To obtain this result, we took advantage of the definition of tensor functions (e.g., $\ln\( \bullet \)$ and $\exp\( \bullet \)$) via series expansion~\cite{higham_2008_functions-of-matrices}.
$\bigo\(\eta^j\)$ is used in the Landau ``big O'' sense, denoting a neglected remainder bounded by a constant multiplied by $\eta^j$.

At the first-order, the incompressibility condition $J = 1$ leads to $\tr{\onegrad} = 0$, which is generally satisfied by linear elastic solutions.
At the second order, the incompressibility condition leads to 
\begin{equation}
     \tr{\twograd} = \frac{1}{2}\tr{\onegrad\onegrad} = \frac{1}{R^{*4}}\( \(\cos\Theta\)^2 - \frac{1}{4} \)
     \, .
     \label{eqn:second_incompress}
\end{equation}

For a boundary point with $R^* = 1$, its distance to the translated center of sphere is maintained to be $a_V$.
Equivalently,
\begin{equation}
    1 + \eta^2\(\frac{2 \, w_R}{a_V} + \frac{\(\sin\Theta\)^2}{4}\) + \bigo\(\eta^3\) = 1
    \quad \Rightarrow \quad
    w_R \bigv_{R^* = 1} = -\frac{a_V}{8} \(\sin\Theta\)^2 
    \, .
    \label{eqn:displ_bc_second}
\end{equation}

To proceed further, we must consider mechanical equilibrium.
For an incompressible neo-Hookean solid, we have Cauchy stress
\begin{equation}
    \cauchy = G \, \leftCG - q \, \iden
    \, ,
\end{equation}
where $\leftCG = \defgrad \, \defgrad^{\transp}$ is the left Cauchy--Green deformation tensor and $q$ is a Lagrange multiplier that can be written in a series expansion form as 
\begin{equation}
    q = G \, \sum_{k = 0}^{\infty} \eta^k \, q_n
    \, .
\end{equation}
At the zeroth order, we require $q_0 = 0$ so that the undeformed body is stress-free.
At the first order, we have $q_1 = -\cos\Theta/R^{*2}$, matching the linear elastic solution.

From our displacement field, we have 
\begin{equation}
    \bB =  \bI + \eta \(\onegrad  + \onegrad^{\top}\) + \eta^2 \(\onegrad\onegrad^{\top}   + \twograd + \twograd^{\top}\) + \bigo\( \eta^3 \)
\, .
\end{equation}

To first order, the vector
\begin{equation}
    \bn = \basis_R + \frac{1}{a_V} \(\eta \, \bv\) + \eta \, \basis_{3} 
       \approx \basis_R - \frac{\eta}{2} \sin\Theta \, \basis_{\Theta}
       \, ,
\end{equation}
is normal to the current cavity surface at a point with reference polar coordinate $\Theta$.
Accordingly, we define a non-normalized tangent vector
\begin{equation}
    \bt = \frac{\eta}{2} \sin\Theta \, \basis_R + \basis_{\Theta}
    \, .
\end{equation}
Our shear-free traction boundary condition for the sphere is then
\begin{equation}
    \(\cauchy \cdot \bn\)\cdot \bt = 0 
    \quad \Rightarrow \quad
    \frac{\epsilon}{2} \sin\Theta \(B_{rr} - B_{\theta\theta}\) + \(1 + \bigo\( \eta^2 \) \)B_{r\theta} = 0
\, .
\end{equation}
To first order, this is equivalent to $H_{r\Theta} + H_{\theta R} = 0$, which is satisfied by our linear elasticity solution.
To second order, we obtain
\begin{equation}
  \(G_{r\Theta} + G_{\theta R}\)\bigv_{R^* = 1} 
   = - \frac{3}{4} \cos\Theta \, \sin\Theta
\, . 
\label{eqn:trac_bc_second}
\end{equation}
In order to satisfy the boundary conditions \cref{eqn:displ_bc_second,eqn:trac_bc_second}, we propose that 
\begin{equation}
    w_R = f\( R \) + g\( R \) \(\cos\Theta\)^2
\, , \quad
    w_{\Theta} = h\( R \)  \cos\Theta \, \sin\Theta
\, .
\end{equation}

Writing the balance of linear momentum in terms of the first Piola--Kirchhoff stress $\pkone = J \, \cauchy \defgrad^{-\transp}$, we have
\begin{equation}
     {\rm Div}\( G \, \defgrad - q \, \defgrad^{-\top} \) = \bzero
\, ,
\end{equation}
where
\begin{equation}
 \defgrad^{-\top} = \bI - \eta \, \bH^{\top} + \eta^2 \( \bH^{\top} \bH^{\top} - \twograd^{\top} \) + \bigo\( \eta^3 \)
\, .
\end{equation}

This leads to 
\begin{equation}
 \Div{
   \eta \, \onegrad + \eta^2 \, \twograd - \(1 - \eta \, \frac{\cos\Theta}{R^{*2}} + \eta^2 \, q_2\(R,\Theta\)\)\(\bI -\eta \, \onegrad^{\top} + \eta^2 \(\onegrad^{\top}\onegrad^{\top} - \twograd^{\top}\)\)
 }
= 0
\, .
\end{equation}
The first-order terms lead to $\Div{ \onegrad + \onegrad^{\top} - \eta \, q_1 \, \bI } = \bzero$, which is satisfied by the linear elastic solution.
The second-order terms lead to
\begin{equation}
   \Div{ \twograd + \twograd^{\top} - \onegrad^{\top}\onegrad^{\top} - \frac{\cos\Theta}{R^{*2}} \onegrad^{\top} - q_2\(R,\Theta\)\bI } = \bzero
\, .
\label{eqn:second_equilibrium}
\end{equation}
This is satisfied if 
\begin{equation}
  q_2 = \alpha\( R \) + \beta\( R \)\(\cos\Theta\)^2
\, ,
\end{equation}
with a form similar to that of $w_R$.

Combining \cref{eqn:second_incompress,eqn:displ_bc_second,eqn:trac_bc_second,eqn:second_equilibrium}, we obtain after some algebraic manipulation,
\begin{equation}
\begin{aligned}
     q_2 &= \frac{1}{5R^{*3}}\(3 \(\cos\Theta\)^2 - 1\)  
\, , \\
     w_R &= \frac{a_V}{40R^{*4}} \( \(4 - 5 \, R^* - 4 R^{* 2}\) - \(12 - 5 \, R^* - 12 R^{* 2}\) \(\cos\Theta\)^2 \)
\, , \\
     w_{\Theta} &= -\frac{a_V}{40 R^{*4}}\(8 - 15R^*\)\cos\Theta \, \sin\Theta
\, .
\label{eqn:second-order-displacement}
\end{aligned}
\end{equation}

% =====================

% =============================================

\newpage
\bibliography{reference} 
\bibliographystyle{unsrtnat}

\end{document}